\documentclass{aa}  
\usepackage{hyperref}
\usepackage{graphicx}
\usepackage[version=3]{mhchem} 
\usepackage{xcolor}
\usepackage{cleveref}
\usepackage{booktabs}
\usepackage{txfonts}
\usepackage{ulem}
\newcounter{chem}
\usepackage{braket}
\newcounter{temp}
\newenvironment{chequation}{%
  \setcounter{temp}{\value{equation}}%
  \setcounter{equation}{\value{chem}}%
}{%
  \setcounter{chem}{\value{equation}}%
  \setcounter{equation}{\value{temp}}%
}
\let\textbf\relax
\begin{document}

   \title{Enhanced formation of interstellar complex organic molecules on carbon monoxide ice}

   \author{G. Molpeceres
          \inst{1,2}
          \and
          K. Furuya
          \inst{1}
          \and
          Y. Aikawa
          \inst{1}
          }

   \institute{
       \textbf{Departamento de Astrofísica Molecular, Instituto de Física Fundamental (IFF-CSIC), C/ Serrano 121, 28006 Madrid, Spain} \\
    \email{german.molpeceres@iff.csic.es}
         \and
   Department of Astronomy, Graduate School of Science, The University of Tokyo, Tokyo 113 0033, Japan\\         
             }

   \date{Received \today; accepted \today}

 
  \abstract
{ We investigate the role of carbon monoxide ice in the chemical evolution of prestellar cores using astrochemical rate equation models. We constrain the ratios of the binding energies on CO ice and \ce{H2O} ice for a series of adsorbates deemed important in diffusive chemistry on \ce{H2O} ices. We later include these ratios in our chemical reaction network model, where the binding and diffusion energies of icy species vary as a function of the surface composition. When the surface coverage of CO increases, the model shows an enhancement of O-bearing complex organic molecules, especially those formed from the intermediate products of CO hydrogenation (e.g. HCO) and \ce{CH3}/\ce{CH2}. Because the binding energy of \ce{CH3}/\ce{CH2} is in the right range, its diffusion rate increases significantly with CO coverage. At $T>$14 K and with less influence, enhanced diffusion of HCO also contributes to the increase of the abundances of COM. We find, however, that chemistry is not always enhanced on CO ice and that the temperature and cosmic ray ionization rate of each astronomical object is crucial for this particular chemistry, revealing a highly non-trivial behavior that needs to be addressed on a per-case basis. Our results are highly relevant in the context of interstellar ice observations with JWST.}


   \maketitle
%

\section{Introduction} \label{sec:introduction}

At the low temperatures of the dense regions of the interstellar medium (10--20 K), and in the absence of any external energy source, all interstellar gas containing heavy elements should be depleted in interstellar dust grains \citep{Willacy1998}. While such a picture is ideal, it highlights the immense importance of interstellar dust grains in the chemical evolution of our cosmos. At the temperatures mentioned above, observations show that dust grains are covered by a thick ice mantle made of different molecules \citep{Boogert2015} such as \ce{H2O}, \ce{CO}, \ce{CO2}, and \ce{CH3OH}, along with minor fractions of the so-called complex organic molecules (COMs), broadly defined as organic molecules with six atoms or more \citep{Herbst2009}. These COMs are fundamental to understanding the evolution of chemical complexity taking place in the very early stages of star and planet formation.

The formation of the ice mantle atop dust grains is sequential. Although \ce{H2O} is formed via surface reactions even in an early stage of molecular clouds \citep{tielens_model_1982, Dulieu2010, ioppolo_laboratory_2008, Lamberts2014, lamberts_water_2013, Meisner2017, Molpeceres2019}, CO is formed in the gas phase and accretes to the grains when the cloud reaches a critical density \citep{Boogert2015}. This phenomenon, known as catastrophic CO freeze-out, is demonstrated by astronomical observations of prestellar cores by an abrupt decay of gas CO abundance \citep{Caselli1999, cazaux_interstellar_2011, taquet_formaldehyde_2012}. In fact, in the traditional picture of COMs formation, the onset of organic chemistry is the hydrogenation of accreted CO to \ce{CH3OH} \citep{Watanabe2002, Fuchs2009, Simons2020, ferrero2023formation}. This picture is still prevalent today, although it is complemented by other theories of COMs formation, such as carbon atom accretion \citep{Molpeceres2021carbon, potapov_new_2021, ferrero2023formation, ferrero_formation_2024} or gas-phase formation \citep{redondo_complex_2017, Cooke2019, ceccarelli_organic_2022}.

One conundrum of interstellar surface chemistry is the impossibility of heavy species roaming the surface of interstellar ice. Although various COMs are detected in cold prestellar cores (e.g. \cite{Bacmann2014, JImenezSerra2016}), diffusive reaction mechanisms such as Langmuir-Hinshelwood are limited by thermal diffusion of the radicals, which can be negligibly low at $\sim$ 10 K. Several alternative mechanisms may explain the presence of COMs in ice mantles and their subsequent return to the gas phase. Some of them are bimolecular reactions through the Eley-Rideal mechanism \citep{le_bourlot_surface_2012, Ruaud2015, ferrero2023formation}, nondiffusive chemistry \citep{Ioppolo2020, Jin2020, Simons2020, Garrod2022} or energetic mechanisms such as UV or e$^{-}$ irradiation \citep{MunozCaro2002, Bennett2006, MunozCaro2014, Turner2018, shingledecker_new_2017, Ishibashi2021}.

The explanation for the emergence of COMs we investigate in this paper is the diffusive chemistry on CO ices. We keep the Langmuir-Hinshelwood mechanism but modify the substrate. On CO ices, the intermolecular forces that bind the adsorbates to the surface are of lower magnitude due to the very low value of the dipole moment of the CO molecule $\left|\mu\right|$=0.122 D \citep{Muenter1975}, in stark contrast to the one of \ce{H2O}, 1.85 D. Lower forces in CO translate into weaker binding and enhanced diffusion of adsorbates on its surface, compared to \ce{H2O} ice. While \citet{Bergin1995} investigated the effect of weaker binding on CO-covered ices on molecular depletion in molecular clouds, in the present work we investigated the effect of enhanced diffusion on the formation of COMs. However, due to its weak binding energy (BE), pure CO ice desorbs thermally at $\sim$18 K \citep{Bisschop2006, Fuchs2006}. Thus, the CO-dominant ice exists only in a narrow temperature range, in which only specific radicals and molecules can diffuse. Nevertheless, laboratory experiments do show rich COM chemistry on CO ice, mainly related to its hydrogenation sequence \citep{Fedoseev2015, fedoseev_simultaneous_2016, Fedoseev2017, Simons2020, Chuang2020}.

In this work, we run astrochemical models to investigate the effect of enhanced diffusion of certain species on CO ice in the aforementioned narrow temperature window. We perform this task using an average proportionality factor $\epsilon_{\mathrm{CO}}$ between BE in CO and \ce{H2O} that is dynamically updated as a function of the dust surface composition. Although most molecules are unaffected by the update of binding energies as a result of low temperature and competition with CO desorption, diffusion of some key radicals is enabled as the CO coverage increases. In \citet{furuyaDiffusionActivationEnergy2022}, we determined that an increase in the diffusion of six radicals, namely \ce{CH3}, \ce{CH2}, \ce{HCO}, \ce{O}, \ce{N}, and \ce{NH} affected the chemistry of cold cores significantly. For these radicals, we derive an individualized scaling factor. We investigated not only the specific influence of CO ice on diffusive chemistry but also how the chemistry couples with other physical magnitudes of the prestellar cores, such as the dust temperature and the cosmic ray ionization rate $\zeta$. Our main aim is to qualitatively show how the surface composition of interstellar ice can modulate molecular abundances in the ISM. This work serves as a stepping stone for the development of models that recognize the heterogeneous composition of interstellar ice.

Our paper is structured as follows. Section \ref{sec:methods} discusses our methodology for astrochemical modeling. We present our results of the grid of astrochemical models of prestellar cores in Section \ref{sec:modelling}, and discuss how the surface composition affects the abundances of certain species. We summarize our main findings in Section \ref{sec:closing} and contextualize our results in the tapestry of modern astrochemical observations.

\section{Astrochemical Models} \label{sec:methods}

We used our in-house astrochemical model {\sc Rokko} described in \citet{furuya_water_2015} and \citet{Furuya2018} using the reaction network and grain surface parameters presented in \citet{Garrod2013} and \citet{bellocheDetectionBranchedAlkyl2014}. Briefly, the \citet{Garrod2013} reaction network comprises almost 1300 species considering gas, surface, and mantle species, coupled through 11,800 reactions. We made several modifications to the network model. The most relevant are setting the BE of C and H atoms to 10,000 K \citep{Wakelam2017, Shimonishi2018, Duflot2021, Molpeceres2021carbon} and 400 K \citep[M06-2X values]{Wakelam2017}, setting the formation of \ce{CO2} through the HOCO radical \citep{molpeceres_cracking_2023}, and expanding the recombination of cations and grains upon collision with negatively charged grains to all cations. For convenience, the BE of \ce{H2} is set equal to the one of H, e.g. 400 K. \textbf{We note that E$^{\ce{CO}}_{\textrm{bin,\ce{H2}}}$ is lower than E$^{\ce{H2O}}_{\textrm{bin,\ce{H2}}}$, as described below, while we set the same BE for H and \ce{H2} (i.e. 400 K) on \ce{H2O} ice.}. Finally, we included the activation energies for radical radical recombinations in \citet{enrique-romero_quantum_2022}. In \citet{enrique-romero_quantum_2022} the authors report activation energies for some radical-radical recombinations on \ce{H2O} ice at different binding sites and in the gas phase. Among other conclusions, the activation energies reported in \citet{enrique-romero_quantum_2022} are binding site dependent. Due to the absence of an H-bond network, in CO ices these activation energies should vanish, as reported by \citet{Lamberts2019} for \ce{CH3CHO}. The applicable activation barriers, when present, will be closer to the gas phase values. Therefore, we have adopted the gas phase values of Table 3 of \citet{enrique-romero_quantum_2022} in our rate equation model.

In order to investigate how chemistry varies with the ice surface composition, the BE of a species is calculated at each time step as follows.

\begin{equation} \label{eq:coverage}
    E_{\text{bin,i}} = \left( 1 - \theta \right) E_{\text{bin,i}}^{\ce{H2O}} + \theta E_{\text{bin,i}}^{\ce{CO}},
\end{equation}

\noindent with $E_{\text{bin}}^{\ce{H2O}}$ the BE of the adsorbate on amorphous solid water, $\theta$ the CO coverage at a given time and $E_{\text{bin}}^{\ce{CO}}$ the CO ice counterpart. Surface coverage, $\theta$, is defined as the fractional abundance of CO on the surface of the ice, which in our models corresponds to 4 monolayers; see below. This approach was used in previous works \citep{Garrod2011, Taquet2014, Furuya2018}. 

In general, $E_{\text{bin,i}}^{\ce{CO}}$ is unknown because the available data is much more scarce, although several studies exist \textbf{\citep{Bisschop2006, Fuchs2006, Fuchs2009, Lamberts2019, mondal_is_2021, sil_chemical_2021, Molpeceres2021a, Ferrari2023, ferrero2023formation}}. \textbf{For pure diffusion energies, the only available comparison is the one between \citet{Hama2012} (on \ce{H2O}) and \citet{Kimura2018} (on CO) for the H atom, finding similar values in the two substrates (See below). } 

To estimate the influence of CO ices in the chemistry of the ISM, we define the binding energies on CO in our model as as:

\begin{equation}  \label{eq:scale}
    E_{\text{bin,i}}^{\ce{CO}} = \epsilon_{\ce{CO}} E_{\text{bin,i}}^{\ce{H2O}}.
\end{equation}

\begin{table}[t]
\begin{center}
\caption{Values of $\epsilon_{\mathrm{CO}}$ (Equation \ref{eq:scale}) and BE of \citet[except for H and \ce{H2}, see text]{Garrod2013} used in our astrochemical models.}
\label{tab:epsilon}
\begin{tabular}{lcc}
\toprule
Adsorbate & $\epsilon_{\mathrm{CO}}$ & BE (K) \\
\bottomrule
\ce{H}$^{a}$  & 1.00 & 400 \\
\ce{H2}$^{b}$       & 0.52 & 400 \\
\ce{CH3}  & 0.39 & 1600 \\
\ce{HCO}  & 0.39 & 2280 \\
\ce{N}  & 0.65 & 720 \\
\ce{O}  & 0.38 & 1320\\
\ce{CH2}  & 0.34 & 1400 \\
\ce{NH}  & 0.36  & 2378 \\
\ce{CO}$^{c}$  & 0.66 & 1300 \\
Rest of surface species$^{b}$ & 0.66 & N/A \\
\bottomrule
\end{tabular}
\tablefoot{$^{a}$: Based on \cite{Wakelam2017}, \cite{Fuchs2009}, \cite{Hama2012} and \cite{Kimura2018}. $^{b}$: See text. $^{c}$: Based on \cite{Furuya2018} and references therein.}
\end{center}
\end{table}

\noindent In the above definition, $\epsilon_{\ce{CO}}$ is the proportionality factor between the binding energies on pure amorphous solid water (ASW) and pure CO ices. In Appendix \ref{sec:appendix1} we show a derivation of $\epsilon_{\mathrm{CO}}$ for the six radicals deemed fundamental for molecular cloud chemistry, i.e. \ce{CH3}, \ce{HCO}, N, O, \ce{CH2}, NH \citep{furuyaDiffusionActivationEnergy2022}, using quantum chemical calculations. \textbf{A particular case is the BE of the H atom, taken from the study of adsorption on a dimer by \citet{Wakelam2017}. In all our models, the BE of H is independent of $\theta$, e.g., it is not updated according to Equation \ref{eq:coverage} based on the similarity between the values found in \citet{Wakelam2017} for \ce{H2O} and in \citet{Fuchs2009} for CO ice, \textbf{also found for diffusion barriers comparing the works of \citet{Hama2012} and \citet{Kimura2018}}. Therefore, we set $E_{\text{bin,H}}^{\ce{CO}}$ = $E_{\text{bin,H}}^{\ce{H2O}}$ = 400 K in our models.} For all the remaining adsorbates not included in the above list (\textbf{and \ce{H2}, due to its importance}), we set an $\epsilon_{\ce{CO}}$=0.66 \citep[][\textbf{and references therein}]{Furuya2018}, \textbf{i.e. the same} as the ratio between the BE of CO on \ce{H2O} ice and CO ice. In Appendix \ref{sec:appendix2}, we show that the selection of a universal $\epsilon_{\ce{CO}}$ for these species does not affect the subsequent conclusions of our calculations. However, we note that 0.66 is already a conservative value and that our quantum chemical estimations suggest that $\epsilon_{\mathrm{CO}}$ should be lower \textbf{for a wide amount of molecules (Appendix \ref{sec:appendix1})}. The values of $\epsilon_{\mathrm{CO}}$ are summarized in Table \ref{tab:epsilon}. In our chemical models, we compared models that enable equation \ref{eq:coverage} (i.e. considering $\epsilon_{\ce{CO}}$) with models where the BE is constant and equivalent to the binding energies on ASW. We track how $\epsilon_{\ce{CO}}$ and $\theta$ affect the evolution of molecules in prestellar cores.

Several nonthermal mechanisms are included in the models. The reactive desorption fraction is given by the scheme of \citet{garrod_non-thermal_2007}:

\begin{align}
    f &= \dfrac{ap}{1 + ap} \label{eq:cd_frac}\\
    p &= \left[ 1 - \dfrac{E_{\text{bin}}}{E_{r}} \right]^{s-1},
\end{align}

\noindent with an $a$ parameter of 0.01, and a reaction energy, $E_{r}$. The parameter $a$ corresponds to the fraction of the reaction energy that is inoculated into the desorption vibration mode. Contrary to thermal diffusion or desorption, which we made dependent on the surface coverage of CO through Equation \ref{eq:coverage}, we did not scale $f$ accordingly.  The reason for this is that the literature does not contain data on $a$ on CO ice. Other nonthermal mechanisms in our model are cosmic-ray induced desorption that follows the Hasegawa-Herbst formulation \citep{hasegawa_three-phase_1993} sampling cosmic ray ionization rates ($\zeta$) in the range between 1$\times$10$^{-18}$--1$\times$10$^{-15}$ s$^{-1}$ (50 logarithmic equispaced values). Photodesorption is included using a constant desorption yield of 1$\times$10$^{-3}$ per photon with a constant extinction coefficient $A_{\rm v}$=100 mag, so CR-induced UV photons dominate the UV radiation field. The gas particle density is set to 2$\times$10$^{5}$ cm$^{-3}$. The temperatures of gas and dust ($T_g$ and $T_d$) are assumed to be the same and are sampled in the range between 8--18 K (50 values). Above that temperature, the thermal desorption of CO becomes important, and the impact of CO coverage on COMs formation is diminished. Combining the sampling of $\zeta$ and $T_d$, our grid of models consists of 2,500 models, that are carried out varying and without varying $\epsilon_{\ce{CO}}$ for comparison, for a total of 5,000 models.

\begin{table}[t]
\begin{center}
\caption{Initial physical conditions in our models. n$_{\text{\ce{H2}}}$ is the \ce{H2} density, A$_v$ the extinction coefficient, $\zeta$ the cosmic-ray ionization rate, $T_g$, $T_{\rm d}$ the gas and dust temperature, respectively. E$_{\text{diff}}$/E$_{\text{bin}}$ is the diffusion to desorption ratio, N$_{\text{act}}$ the number of chemically active layers, and $a$ corresponds to the energy conversion parameter in the chemical desorption scheme of \cite{garrod_non-thermal_2007}.}
\label{tab:phys}
\begin{tabular}{cc}
\toprule
Parameter & Value \\
\bottomrule
n$_{\text{\ce{H2}}}$   &  1$\times$10$^{5}$ cm$^{-3}$ \\ 
$A_{\rm v}$ & 100 mag \\
$\zeta$ & \{ 1$\times$10$^{-18}$--1$\times$10$^{-15}$ \} s$^{-1}$ \\
$T_{\rm g}$, $T_{\rm d}$ & \{ 8--18 \} K \\
E$_{\text{diff}}$/E$_{\text{bin}}$ & 0.4 \\
N$_{\text{act}}$ & 4 \\
$a$ & 0.01 \\
\bottomrule
\end{tabular}
\end{center}
\end{table}

Following the previous models of the gas-grain reaction network, the relationship between $E_{\text{bin}}$ and $E_{\text{diff}}$ (the diffusion energy) is assumed to be $E_{\text{diff}}$/$E_{\text{bin}}$ = 0.4, e.g. a constant value. The drawbacks of this approach are discussed in \citet{furuyaDiffusionActivationEnergy2022}, but this choice allows us to keep the parameter space search (e.g. $\zeta$ and $T_d$) tractable. While quantitative conclusions are indeed affected by a different choice of $E_{\text{diff}}$/$E_{\text{bin}}$, our qualitative conclusions about the enhancement of the chemistry in CO are not affected by this model choice as we show in Appendix \ref{sec:appendix3}. We consider the reaction-diffusion competition scheme in all our models \citep{changGasgrainChemistryCold2007}.  Finally, in our models, we set the four topmost (N$_{\text{act}}$=4) layers as chemically active and regarded as surface species. On the contrary, the mantle chemistry below these four monolayers is considered inactive. A summary of the physical parameters of our simulations is presented in Table \ref{tab:phys}.

\begin{table}[t]
\begin{center}
\caption{Initial species and elemental abundances employed in our chemical models. Abundances are  with respect to H nuclei. \textbf{A(B) notation is used to denote A$\times$10$^{B}$.}}
\label{tab:abundances}
\begin{tabular}{lc||lc}
\toprule
Element & Abundance & Species & Initial abundance \\
\bottomrule
H & 1.00(0) & \ce{H2} & 5.00(-1) \\
\ce{He} & 9.75(-2) & \ce{He} & 9.75(-2) \\
\ce{N}  & 2.47(-5) & \ce{N}  & 2.47(-5) \\
C & 7.86(-5) & \ce{CO} & 5.89(-5) \\
O & 1.80(-4) & \ce{C+} & 1.97(-5) \\ 
S & 9.14(-8) & \ce{H2O}$^{a}$ & 1.21(-4) \\
P & 1.00(-9) & \ce{S+} & 9.14(-8)  \\
\ce{Si} & 9.74(-9) & \ce{P+} & 1.00(-9)  \\
Fe & 2.74(-9) & \ce{Si+} & 9.74(-9) \\
Na & 2.25(-9) & \ce{Fe+} & 2.74(-9) \\
Mg & 1.09(-8) & \ce{Na+} & 2.25(-9) \\
Cl & 2.16(-10) & \ce{Mg+} & 1.09(-8) \\
 & & \ce{Cl+} & 2.16(-10) \\
\bottomrule
\end{tabular}
\tablefoot{$^{a}$: On ice.}
\end{center}
\end{table}

\begin{figure}
    \centering
    \includegraphics[width=\linewidth]{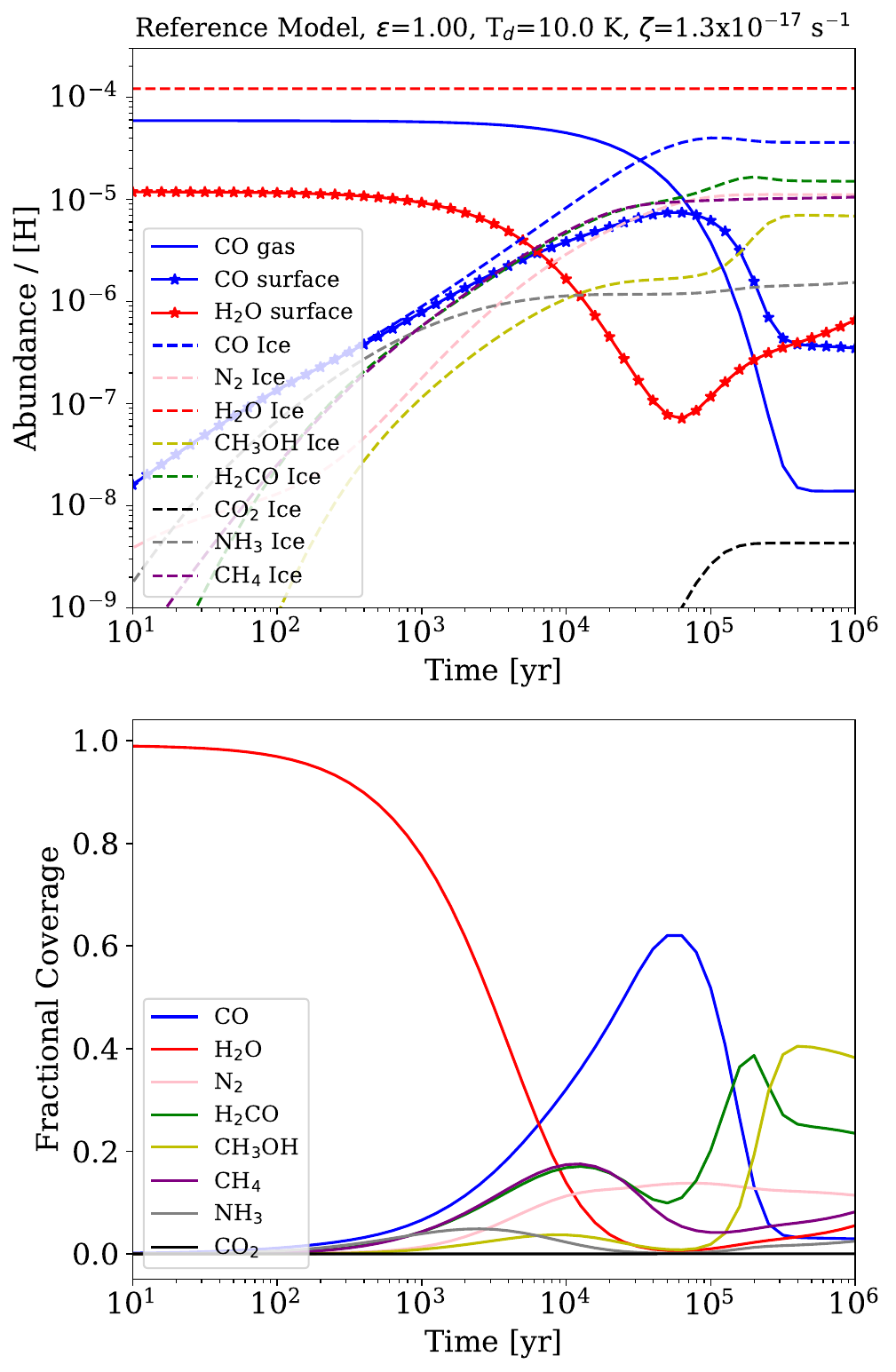} \\
    \caption{Top - Evolution of the abundances of main ice species, surface \ce{H2O} and CO, and gas CO in our reference model. Bottom - The fractional surface coverage of the main ice species corresponds to the fraction of molecules in the four topmost layers of a dust grain. }
    \label{fig:fiducial}
\end{figure}

The initial conditions of the models are selected to maximize the impact of CO chemistry; we start with a significant fraction of C and O nuclei locked in gas-phase CO (before freeze-out) and \ce{H2O} ice. The elemental abundances and the initial molecular/atomic abundances are presented in Table \ref{tab:abundances}. The elemental abundances are taken from \citet{Aikawa1999}, low-metallicity values. We consider a single population of interstellar grains of size 0.1~$\mathrm{\mu m}$ with approximately 10$^{6}$ binding sites per grain. 

We define our reference model with parameters of $\epsilon_{\mathrm{CO}}$ value of 1.0, $T_{\rm d}$ of 10.0 K and $\zeta$ of 1.3$\times$10$^{-17}$ s$^{-1}$.  The temporal variations of molecular abundances are shown in Figure \ref{fig:fiducial} for the main ice species. In the bottom panel of Figure \ref{fig:fiducial}, we show the fractional surface coverage of the main ice species. In the model, we observe a clear and abrupt depletion of CO gas around 5$\times$10$^{4}$ yr and the formation of an apolar-rich ice surface in a time window between 4$\times$10$^{4}$ and 3$\times$10$^{5}$ yr. Thus, in section \ref{sec:modelling}, we present molecular abundances at 2$\times$10$^{5}$ yr for a grid of models. This time corresponds to an average lifetime of a pre-stellar core \citep{andre2014filamentary} and also a time after most of the catastrophic CO freeze-out has taken place when the effects of the CO chemistry are more apparent. The surface coverage of CO ($\theta$) peaks at 6$\times$10$^{4}$ years, $\theta\sim$0.65, with \ce{CH4}, \ce{N2} and \ce{H2CO} as other main components. Because the freeze-out time scale depends only slightly on the temperature at 10-20 K, the choice of 2$\times$10$^{5}$ yr is also justified for the rest of the surface temperatures. 

\section{Results} \label{sec:modelling}

\subsection{Initial analysis} \label{sec:initial_analysis}

\subsubsection{Diffusion times as a function of time and $T_{d}$}

\begin{figure}
    \centering        
    \includegraphics[width=\linewidth]{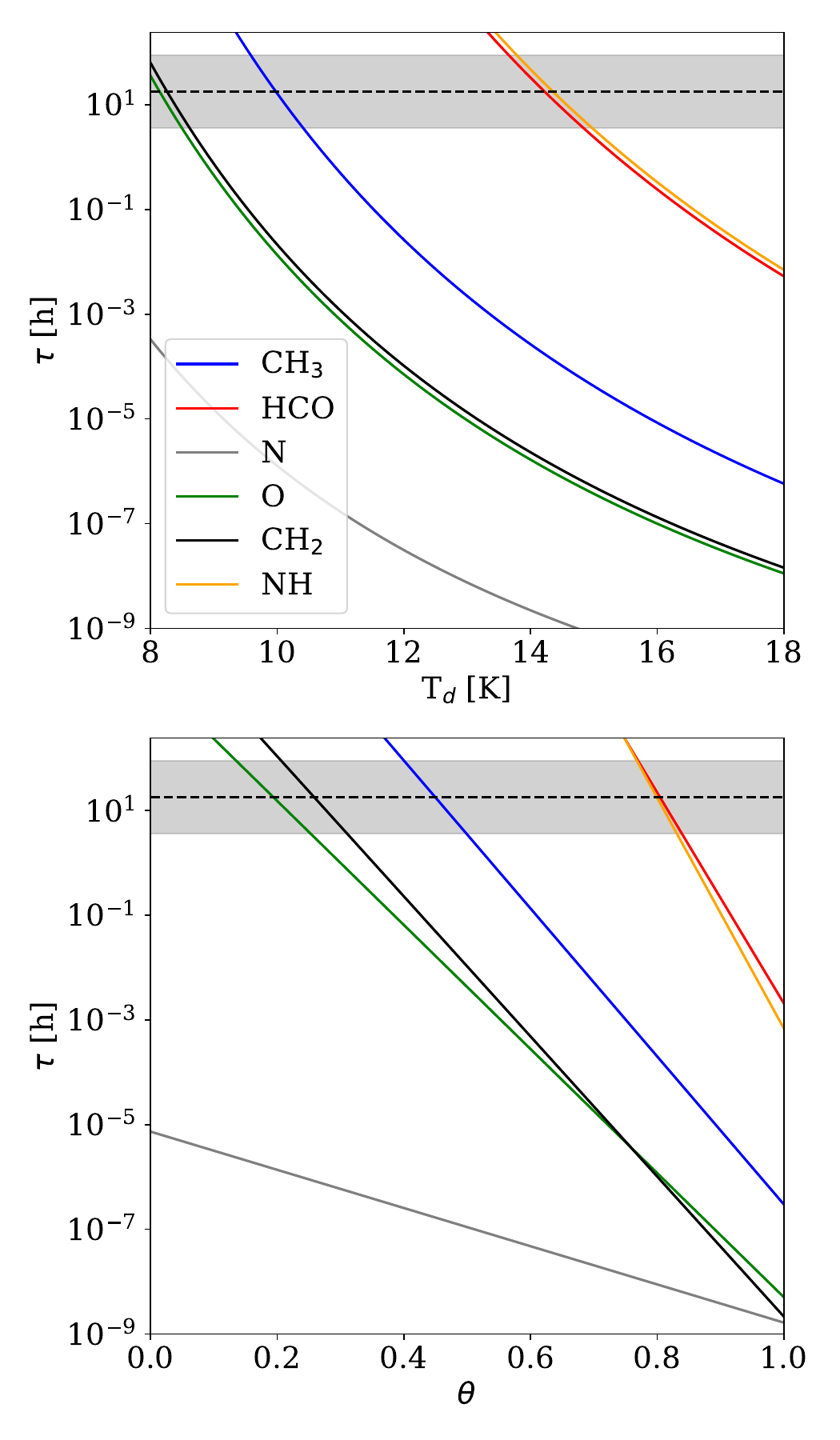} \\
    \caption{Top - \textbf{Diffusion timescale} as a function of the dust temperature ($T_\text{d}$) for $\theta$=0.65  Bottom - as a function of the surface coverage of CO ($\theta$) at 12 K for the main radicals considered \citep{furuyaDiffusionActivationEnergy2022}. The dotted black line and the shaded region around it show the H accretion timescale for different $n_{H_{\text{atom}}}$ (see Section \ref{sec:initial_analysis}).  }
    \label{fig:hopping}
\end{figure}

Qualitative differences in the chemistry on \ce{H2O}-dominant ices and CO-dominant ices would be caused by species whose diffusion rate is negligible on \ce{H2O} ice but significant on CO ice. Hence it is beneficial to check the diffusion timescale ($\tau$, i.e. inverse of the rate constant for diffusion):

\begin{equation} \label{eq:hopping}
    \tau = 1 /\left(\nu \exp \left[\dfrac{-0.4\left(\left( 1 - \theta \right) E_{\text{bin,i}}^{\ce{H2O}} + \theta E_{\text{bin,i}}^{\ce{CO}}\right)}{T_d}\right]\right),
\end{equation}

\noindent of the key radicals considered in this work on pure \ce{H2O} ice and pure CO ice (Figure \ref{fig:hopping}), as a function of $T_{d}$ and $\theta_{\text{CO}}$. In equation \ref{eq:hopping} we assume an $E_{\text{diff}}$/$E_{\text{bin}}$ ratio of 0.4 and a preexponential factor of diffusion ($\nu$) of 1$\times$10$^{12}$ s$^{-1}$. $\tau$ must be confronted against the competitive process that is the hydrogenation of the radicals, whose timescale depends on the H accretion (and diffusion rate). The H accretion frequency can be determined as

\begin{equation}
    f_{H} = \dfrac{1}{\sigma \braket{v_{H}} n_{H_{\text{atom}}} }
\end{equation}

\noindent where $\sigma$ is the cross-section of a dust grain, $\braket{v_{H}}$ is the average thermal velocity of the H atom at 10 K and $n_{H_{\text{atom}}}$ is the number density of H atoms ($\sim$1 cm$^{-3}$ for $\zeta$=1.3$\times$10$^{-17}$ s$^{-1}$, \cite{Goldsmith2005}). This timescale is depicted in Figure \ref{fig:hopping} as a dotted black line. Because $f_{H}$ depends strongly on $\zeta$, we consider $n_{H_{\text{atom}}}$ in a region that spans a factor 5 from the above value (represented as a shaded gray box in Figure \ref{fig:hopping}).

Figure \ref{fig:hopping} (Top panel) indicates that below 12 K and in a $\theta$=0.65 CO mixed ice, only species with effective $E_{\text{diff}}$ below 800 K may experience rapid diffusion, for example N, O or \ce{CH2}. We note that mobility of radicals such as \ce{CH2} or O is not negligible even below 10 K. Slow but plausible diffusion of species with binding energies close to 1500 K is also enabled (i.e. \ce{CH3}) close to 12 K. Above 14 K, the diffusion of molecules with binding energies around 1500 K is fast. Between 12--14 K, radicals with binding energies higher than 2000 K start to diffuse slowly (HCO, NH). The bottom panel of Figure \ref{fig:hopping} is informative to rationalize a ``critical" CO coverage at which the chemistry on CO is truly enabled. This surface coverage is roughly 0.60, which is close to the maximum $\sim$0.65 found in our reference model (Figure \ref{fig:fiducial}, bottom panel). We note that owing to the differences in physical conditions within the grid of models to be presented in this section, the maximum $\theta$ can vary between models; see below. However, a valuable piece of information derived from this analysis is the threshold concentration of CO molecules on the surface to activate new chemistry. 

In terms of chemistry, including a coverage-dependent term in the diffusion allows for rapid diffusion of molecules below 12 K, as mentioned above. Therefore, the abundance of molecules whose parent species are \ce{CH3} (and \ce{CH2} as the parent of \ce{CH3}), or O and N atoms, are susceptible to their formation of the apolar ice layer. Above 12 K, \ce{HCO} or \ce{NH} also begin to diffuse, competing with the rapid consumption of CO and the consequential increase in E$_{\text{bin}}$. Therefore, mainly due to the delicate balance established by CO consumption through several mechanisms (hydrogenation, desorption), the species with the most likely impact on the chemistry of apolar ices at very low temperatures are \ce{CH3}, \ce{CH2}, N and O because their diffusion rate is significant considering significant CO ice coverage while they are negligible with $\theta$=0. 

In the following sections, we present the results of our chemical model for some selected species and discuss how the CO surface coverage affects these specific molecules. We discuss the chemical reactions that are enabled on CO ice and that lead to the different abundances of the target species. The selection of specific molecules affected by the different chemistry on CO ice does not imply that these are the only molecules susceptible to enhanced chemistry on apolar ice. However, selecting key molecules presented in the extensive \citet{Garrod2013} chemical network is necessary to keep our discussion as general as possible.

\subsubsection{The importance of $\zeta$} \label{sec:importance}

As described in Section \ref{sec:methods} and Table \ref{tab:phys}, we vary two physical parameters in our grid of models: surface temperature, $T_{d}$, and cosmic ray ionization rate, $\zeta$. Although the effect of $T_{d}$ is evident from Equation \ref{eq:hopping}, the importance of $\zeta$ may not be obvious. In fact, $\zeta$ is the most important factor affecting CO coverage and CO-related chemistry on the ice surface.

At low temperatures, CO is one of the few abundant closed shell species that can be chemically converted through the chain of (forward and backward) reactions \citep{Watanabe2002, Fuchs2009}:

\begin{chequation}
    \begin{align} \label{eq:CO}
   \ce{CO &->[+\text{H}] HCO ->[+\text{H}] H2CO ->[+\text{H}] CH3O ->[+\text{H}] CH3OH}\\
   \ce{CH3OH &->[+\text{H},-\ce{H2}] CH2OH ->[+\text{H},-\ce{H2}] H2CO ->[+\text{H},-\ce{H2}] HCO ->[+\text{H},-\ce{H2}] CO  }
\end{align} 
\end{chequation}

\noindent Therefore, the abundance of CO on the surface is inversely correlated with the presence of H atoms. The steady-state H abundance is determined by $\zeta$ \citep{Goldsmith2005}, because cosmic rays dissociate \ce{H2} molecules very effectively into two H atoms that later accrete on the ice and reduce the abundance of CO. In summary, in models with higher $\zeta$, CO-coverage is lower, reducing COM formation through efficient diffusion on apolar ices. This fact is slightly counterintuitive because higher $\zeta$ is known to produce more radicals on \ce{H2O} due to the fragmentation of molecules. However, in CO ices, removing the CO layer by H atoms overwhelms the latter effect. As we show in the next section \ce{CH3OH} is an exception to this rule, precisely because it is the endpoint of the CO hydrogenation sequence.

The other magnitude that depends on $\zeta$ is the cosmic-ray-induced desorption that returns molecules to the gas phase via stochastic heating after the collision \citep{hasegawa_three-phase_1993}. This effect indirectly depends on the binding energy and synergizes with chemical desorption. It will be briefly discussed in Section \ref{sec:gas}. However, we found it to be less critical in the context of the chemistry of CO ices than in the production of H atoms discussed above.

\subsection{Chemistry on CO activated at $\leq$ 12 K} \label{sec:modelCOCH3}

We start our analysis with an examination of the abundances of molecules originating from diffusion below 12 K (Figure \ref{fig:hopping}). As we introduced in Section \ref{sec:modelling}, our analysis of molecular abundances is performed at 2$\times$10$^{5}$ yr. Disentangling the influence of surface composition requires careful examination of many factors that influence surface chemistry in interstellar environments. These include not only the binding energies updated by the chemical composition of the surface but also how this effect couples with the cosmic ray-induced processes. It is customary to mention that $E_{\text{diff}}$/$E_{\text{bin}}$ is an important factor in the model \citep{furuyaDiffusionActivationEnergy2022}. We assume a constant value for all of our simulations. This is a caveat of our models. However, we note that we can estimate at least qualitatively which molecules and radicals would play a critical role in the CO-dominated ice as described in Section \ref{sec:initial_analysis} when a different ratio of $E_{\text{diff}}$/$E_{\text{bin}}$ is adopted (Appendix \ref{sec:appendix3}).

\begin{figure*}
    \centering
    \includegraphics[width=0.45\linewidth]{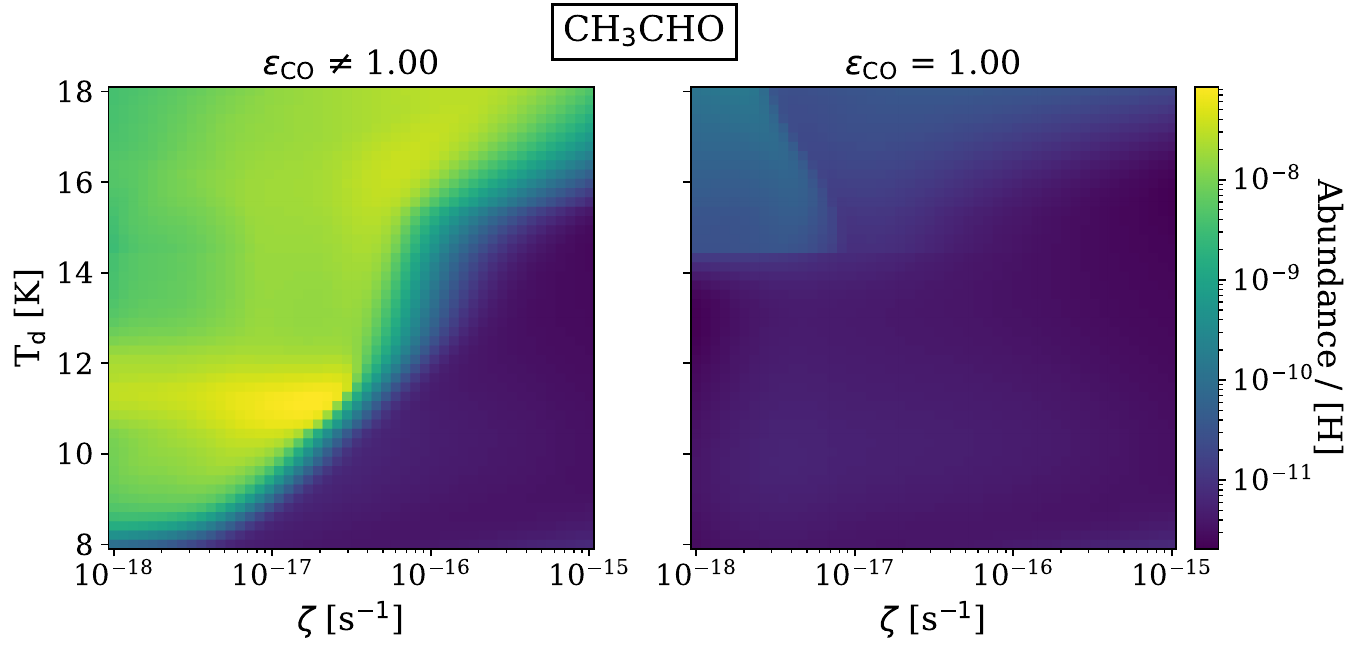} 
    \includegraphics[width=0.45\linewidth]{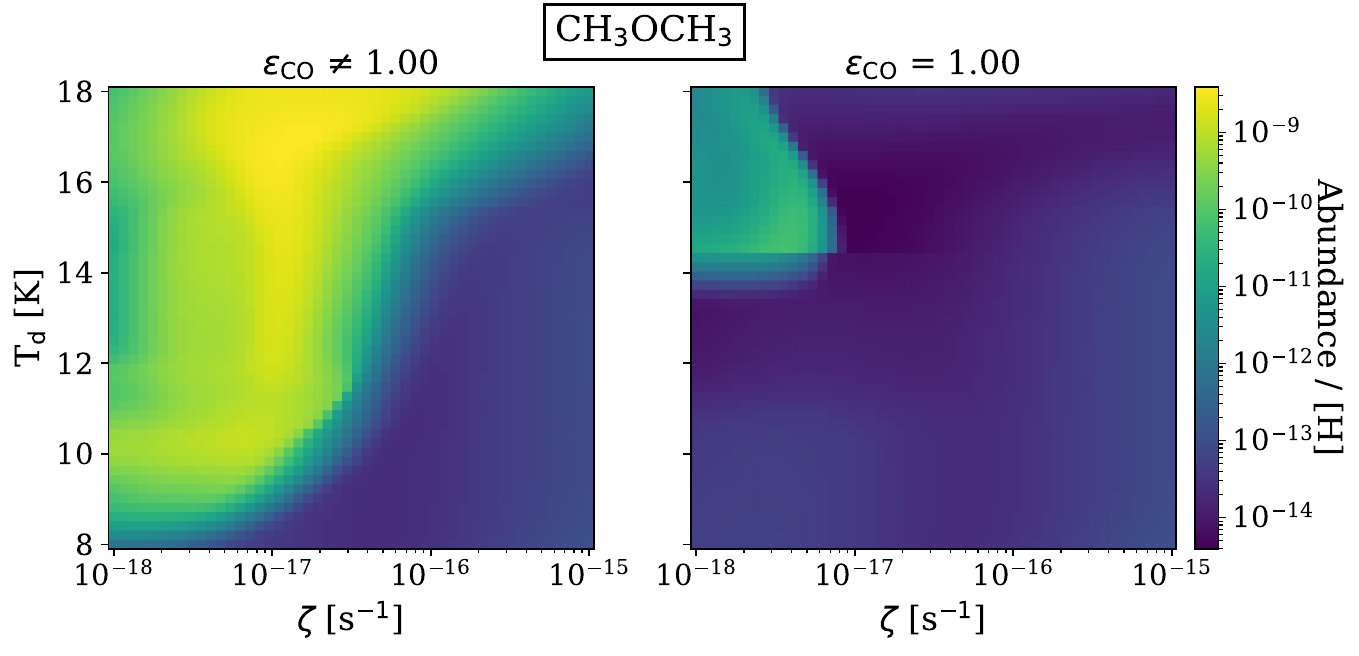} \\ 
    \includegraphics[width=0.45\linewidth]{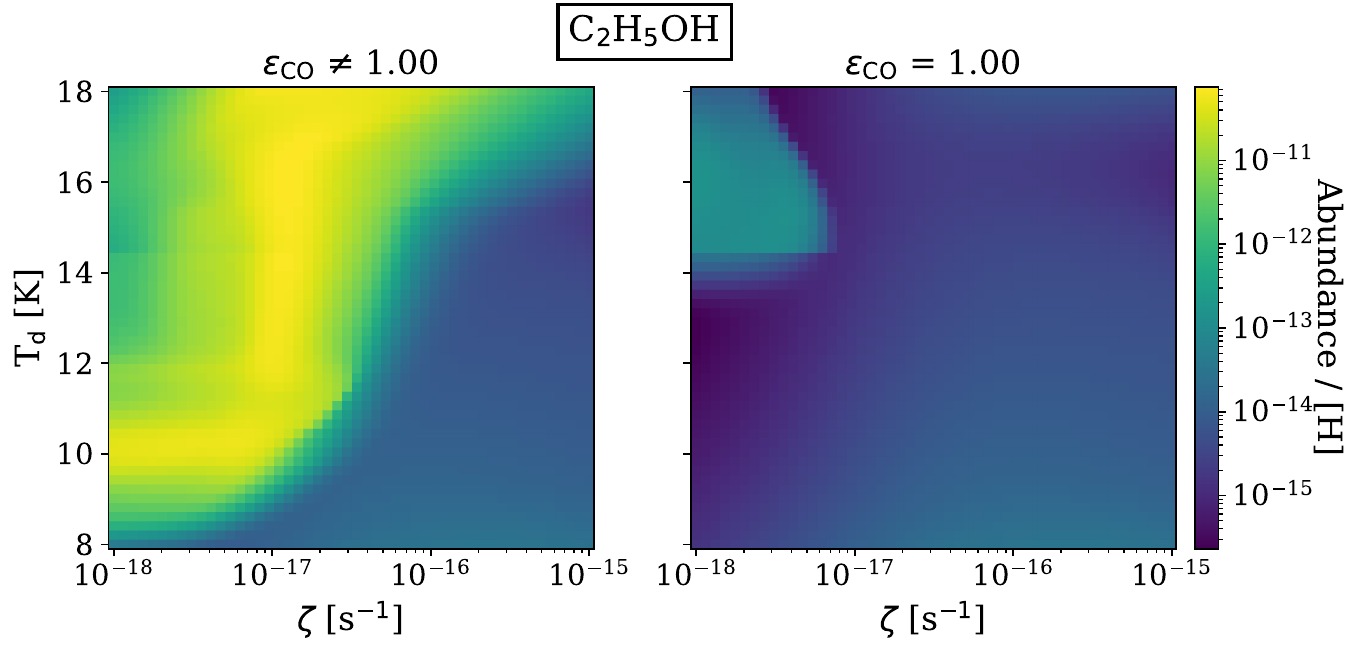} 
    \includegraphics[width=0.45\linewidth]{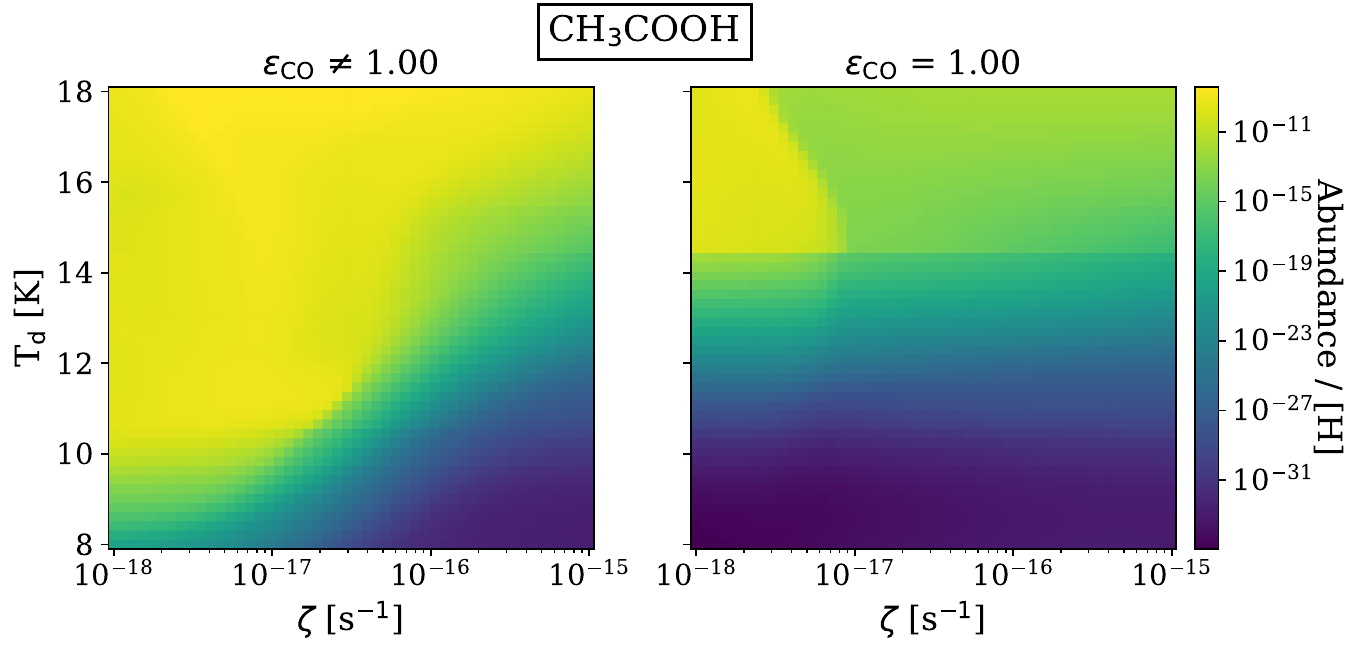} \\
    \includegraphics[width=0.45\linewidth]{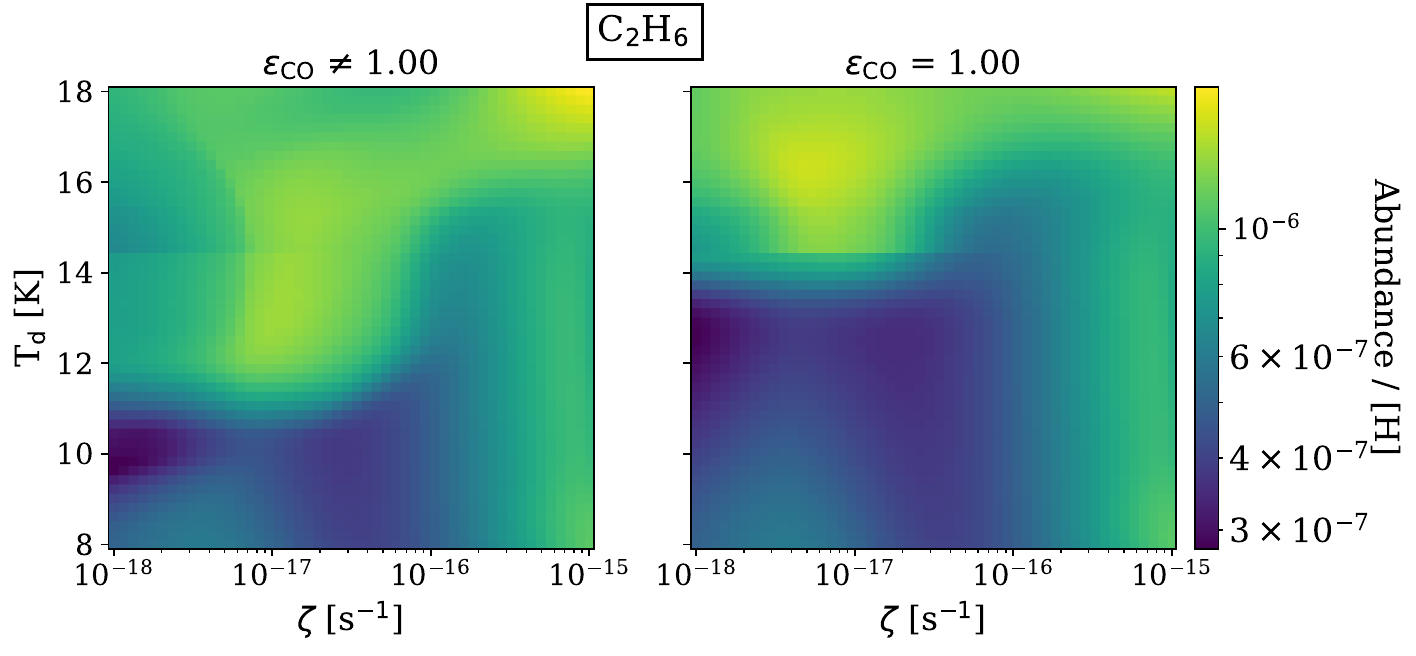} 
    \includegraphics[width=0.45\linewidth]{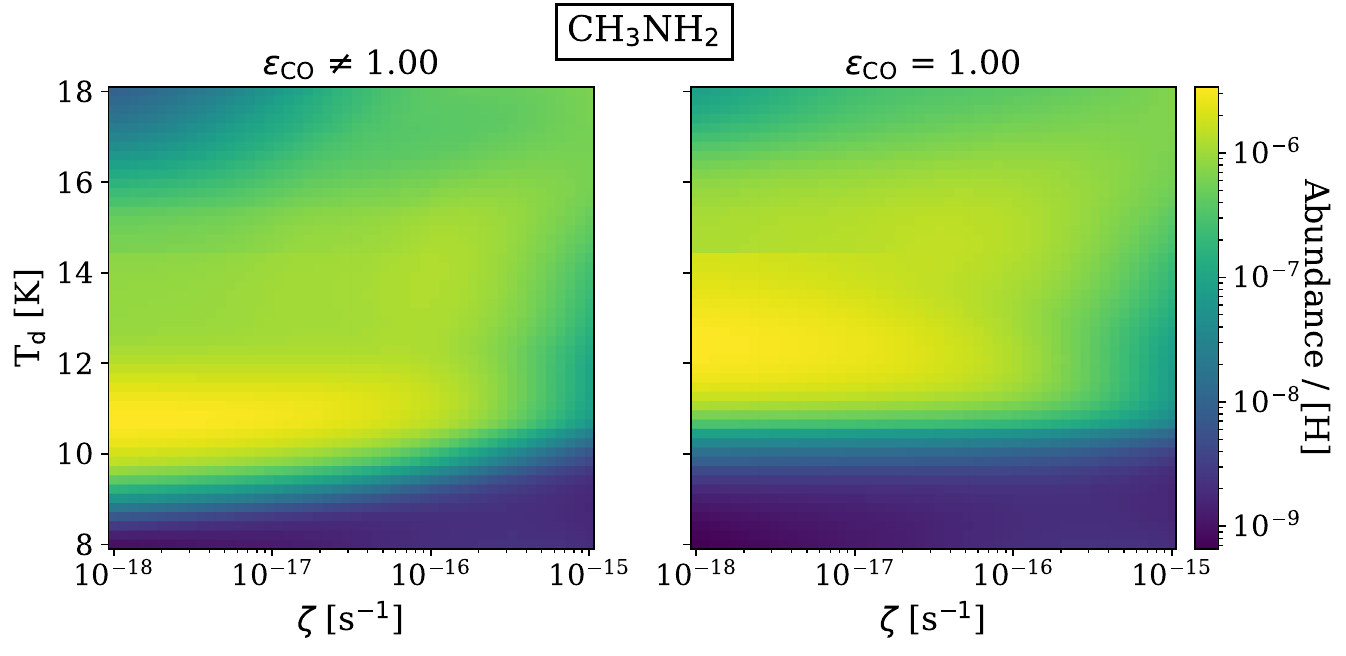} 
    \caption{\textbf{Ice abundances at t=2$\times$10$^5$ yrs in models of various $\zeta$ and $T_{\rm d}$. The selected molecules are \ce{CH3CHO}, \ce{CH3OCH3}, \ce{C2H5OH}, \ce{CH3COOH}, \ce{C2H6}, and \ce{CH3NH2}.}}
    \label{fig:heatmaps:12K}
\end{figure*}

Figure \ref{fig:heatmaps:12K} shows the abundances of assorted molecules at $t$=2$\times$10$^{5}$ yr as a function of $\zeta$ and $T_\text{d}$ for models of $\epsilon_{\ce{CO}}$ $\neq$ 1.0 and $\epsilon_{\ce{CO}}$=1.0. The latter corresponds to the classical model, in which the adsorption and diffusion energies do not depend on the surface composition of the ice. We selected molecules directly related to the radicals analyzed in Section \ref{sec:initial_analysis} whose diffusion timescale is significantly affected by the composition of the surface ice. In the following, we highlight \ce{CH3CHO}, \ce{CH3OCH3}, \ce{C2H5OH}, \ce{CH3COOH}, \ce{C2H6}, and \ce{CH3NH2}. Although the first four molecules are sensitively dependent on $\epsilon_{\ce{CO}}$, the last two species are less affected by a variety of factors that we detail in the following paragraphs. Other molecules, such as \ce{CH3OH}, \ce{CH3CN}, or \ce{C2H4}, are also described as examples of unaffected molecules.

\paragraph{\ce{CH3CHO} (Acetaldehyde)}

Acetaldehyde is the molecule most affected by the update of the binding energies. Introducing the $\theta$ dependence in our models can increase the abundance of \ce{CH3CHO} up to four orders of magnitude in the temperature range investigated in the present work. There are three related factors.

First, diffusion of \ce{CH3} and \ce{CH2} on a CO-dominated surface enables the surface reactions:

\begin{chequation}
\begin{align} 
    \ce{CH3 + HCO &-> CH3CHO} \label{eq:CH3CHO} \\
    \ce{CH2 + HCO &-> CH2CHO ->[\text{H}] CH3CHO}
\end{align}
    
\end{chequation}

\noindent Reaction \ref{eq:CH3CHO} was investigated on pure \ce{H2O} ice by \cite{enrique-romero_theoretical_2021}, finding a critical dependence on the $E_{\text{diff}}$/$E_{\text{bin}}$, with respect to the competitive channel:

\begin{chequation}
\begin{equation} \label{eq:CH4CO}
    \ce{CH3 + HCO -> CH4 + CO}.
\end{equation}
\end{chequation}

\noindent \cite{enrique-romero_theoretical_2021} find a dominance of abstraction (Reaction \ref{eq:CH4CO}) over addition (Reaction \ref{eq:CH3CHO}) for the reaction at a cavity on \ce{H2O} ice with $E_{\textrm{diff}}$/$E_{\textrm{bin}}$=0.3, due to the interactions of the adsorbates with the \ce{H2O} substrate in a particular binding mode. This dominance vanishes for E$_{\text{diff}}$/E$_{\text{bin}}$=0.4. In general, the competition between addition and abstraction on \ce{H2O} ice, not only for \ce{CH3CHO} but for many COMs, depends on the binding site \citep{Enrique-Romero2016, Butscher2017, enrique-romero_quantum_2022}. Mixed interstellar ice or CO ices have weaker binding sites than pure \ce{H2O} ice so the contribution of the particular binding mode becomes less significant. Since both reactions \ref{eq:CH3CHO} and \ref{eq:CH4CO} do not have an intrinsic activation barrier (i.e., they are barrierless in the gas phase), we expect an equal branching ratio, i.e. 0.5 each. The lack of intrinsic barriers was also found by \citet{Lamberts2019} for the formation of \ce{CH3CHO} on the CO ice. Although \citet{enrique-romero_theoretical_2021} relies on the uncertainty of the diffusion energies of \ce{CH3} and \ce{HCO} to postulate the possibility of formation of \ce{CH3CHO}, we consider that the surface coverage of CO plays an active role in promoting the reaction. Our conclusions are complementary and have different applicability, depending on the evolutionary stage of the molecular cloud, e.g. the \ce{CO}/\ce{H2O} ratio on the ice surface. 

The second reason for the enhanced formation of \ce{CH3CHO} is the abundance of HCO on the surface of the grains. Compared to other radicals, HCO is relatively abundant thanks to CO hydrogenation (Equation \ref{eq:CO}). 

The third reason is that, at slightly higher temperatures, HCO also starts to be relatively mobile, enhancing the reaction \ref{eq:CH3CHO}. All these three factors allow the formation of acetaldehyde at low and intermediate $\zeta$, in the entire temperature range considered in this study.

\paragraph{\ce{CH3OCH3} (Dimethyl ether)}

The arguments presented for \ce{CH3CHO} can be directly transferred to \ce{CH3OCH3}. Therefore, we observe a remarkable similarity between the graphs corresponding to these molecules in Figure \ref{fig:heatmaps:12K}. The difference between these two species is the second radical that participates in the reaction, that for \ce{CH3OCH3} is \ce{CH3O}:

\begin{chequation}
\begin{equation}
    \ce{CH3 + CH3O -> CH3OCH3},
\end{equation}
\end{chequation}

\noindent with the competing channel with an activation barrier of 204 K \citep{enrique-romero_quantum_2022}:

\begin{chequation}
\textbf{\begin{equation}
    \ce{CH3 + CH3O -> CH4 + H2CO}.
\end{equation}}
\end{chequation}

\noindent Therefore, the lower abundance of \ce{CH3OCH3} to \ce{CH3CHO} is due to the lower abundance of its parent radical \ce{CH3O} than HCO, about one order of magnitude. The higher BE of \ce{CH3O} also makes it effectively immobile, unlike what happens with HCO, reducing the relative abundance with respect to \ce{CH3CHO} at higher T$_{d}$. Both \ce{HCO} and \ce{CH3O} are formed through the CO hydrogenation sequence \citep{Watanabe2002, Fuchs2009, Simons2020}.

\paragraph{\ce{C2H5OH} (Ethanol)}

The set of the three molecules most affected by \ce{CH3} interacting with hydrogenated CO products concludes with ethanol, \ce{C2H5OH}. Like in the two previous molecules, the enhanced reaction on CO ice is related to the hydrogenation sequence on CO ice (Equation \ref{eq:CO}). The reaction that enhances the formation of \ce{C2H5OH} is:

\begin{chequation}
\begin{equation}
    \ce{CH3 + CH2OH -> C2H5OH},
\end{equation}
\end{chequation}

\noindent that again competes with the following reaction with an 1170 K barrier:

\begin{chequation}
\begin{equation}
    \ce{CH3 + CH2OH -> CH4 + H2CO}.
\end{equation}
\end{chequation}

\noindent The very similar profiles found for \ce{CH3CHO}, \ce{CH3OCH3}, and \ce{C2H5OH} in Figure \ref{fig:heatmaps:12K} confirm this explanation, and, as discussed for \ce{CH3OCH3}, variations between them can be attributed to variations in the abundance of the parent species.

\paragraph{\ce{CH3COOH} (Acetic Acid)}

Acetic acid abundance is greatly enhanced by CO ices. In contrast with the ternary of previous molecules, the reason for its increase is due to the enabling of the reaction

\begin{chequation}
\begin{equation} \label{eq:hoco} 
    \ce{CO + OH -> HOCO},
\end{equation}
\end{chequation}

\noindent following the recent study of \cite{molpeceres_cracking_2023}. Reaction \ref{eq:hoco} was previously considered to form directly \ce{CO2} \citep{ioppolo_surface_2011, Garrod2011,oba_experimental_2010, terwisscha_van_scheltinga_formation_2022}, but in our recent study we showed otherwise. Hence, the chemistry of HOCO can be initiated on the surface of the CO-rich ice. Associated with the selective production of HOCO, we also find an increase in the diffusion of CO ($\epsilon$=0.66) in mixed ices ($\theta \leq$ 0.5) and self-diffusion in CO-rich ices ($\theta$\textgreater0.5), further increasing the production of HOCO. With a high concentration of HOCO radicals on the surface, the formation of \ce{CH3COOH} can proceed through:

\begin{chequation}
\begin{equation} \label{eq:ch3cooh} 
    \ce{HOCO + CH3 -> CH3COOH}.
\end{equation}
\end{chequation}

\noindent Both reactions \ref{eq:hoco} and \ref{eq:ch3cooh} are facile on CO ice in our models, in contrast to \ce{H2O} ice, which requires a threshold temperature of ($\sim$ 12 K) for the diffusion of CO \citep{Garrod2011}. On CO ices, the lack of this threshold summed to the easy diffusion of \ce{CH3} is what causes the dramatic increase in the production of this molecule. However, we note the subtle differences in the HOCO formation barriers on \ce{H2O} and CO ices through reaction \ref{eq:hoco} \citep{molpeceres_cracking_2023}. Regretfully, we can not capture these differences with our models yet. 

\paragraph{\ce{C2H6} (Ethane)}

Ethane is an interesting molecule in this context because it presents three different regimes in its abundance heat maps in Figure \ref{fig:heatmaps:12K}. The formation of ethane on CO ice is enhanced by the surface reaction:

\begin{chequation}
    \begin{equation} \label{eq:ch3ch3}
    \ce{CH3 + CH3 -> C2H6},
\end{equation}
\end{chequation}

\noindent primarily around 12 K with deviations between $\epsilon_{\ce{CO}}$=1.0 and $\epsilon_{\ce{CO}}\neq$1.0 of roughly a factor 3. However, above 12 K, we observe an increase in the production of \ce{C2H6} that cannot be explained by the recombination of \ce{CH3} alone. A careful inspection of the reaction rates shows that the reaction:

\begin{chequation}
    \begin{equation} \label{eq:c2h5hco}
    \ce{C2H5 + HCO -> C2H6 + CO},
\end{equation}
\end{chequation}

\noindent is responsible for the increase in \ce{C2H6} abundance at medium temperatures (and medium $\zeta$), promoted by the diffusion of \ce{HCO} above 12 K. The less marked influence of the CO ice for this molecule is due to the importance of the reaction

\begin{chequation}
    \begin{equation}
    \ce{C2H5 + H -> C2H6}
\end{equation}
\end{chequation}

\noindent across our parameter space, which is the last step in the hydrogenation of \ce{C2H2} \citep{Kobayashi2017,molpeceres_radical_2022}. Although reactions \ref{eq:ch3ch3} and \ref{eq:c2h5hco} do enhance the formation of \ce{C2H6}, they are secondary channels. In this study, we found that, when a molecule can be formed by hydrogenation of abundant species, the influence of CO ice chemistry is minor or null and only molecules without a clear hydrogenation route have a significant increase in abundance (at low $\zeta$).

\paragraph{\ce{CH3NH2} (Methylamine)}

In Figure \ref{fig:heatmaps:12K}, we observe that the models for $\epsilon_{\ce{CO}}\neq$1.0 and $\epsilon_{\ce{CO}}$=1.0 present a similar profile but are slightly vertically displaced, meaning that the same chemistry begins to operate, but at a lower temperature for $\epsilon_{\ce{CO}}\neq$1.0. Reactions that are common to the $\epsilon_{\ce{CO}}\neq$1.0 and $\epsilon_{\ce{CO}}$=1.0 models are

\begin{chequation}
  \begin{align}
    \ce{CH3NH + H &-> CH3NH2} \\
    \ce{CH2NH2 + H &-> CH3NH2}.
\end{align}  
\end{chequation}

\noindent The enhanced formation of \ce{CH3NH2} at low temperatures on CO ice is a consequence of the following reactions:

\begin{chequation}
\begin{align}
    \ce{CH3 + N  &-> CH2NH} \\
    \ce{CH3 + NH &-> CH3NH} \\
    \ce{CH2 + NH2 &-> CH2NH2} \\
    \ce{CH2 + NH &-> CH2NH} \\
    \ce{CH3 + NH2 &-> CH3NH2} 
\end{align}    
\end{chequation}

\noindent The first four products are further hydrogenated to \ce{CH3NH2}. The heatmap in figure \ref{fig:heatmaps:12K} is vertically displaced because the diffusion of N atom is activated early on water ice and CO ice only plays a role at a very low temperature. At higher $\zeta$, \ce{NH3} formation is favoured, and \ce{NH$_{x}$} (x$\leq$2) abundance is lower on the surface. At high $T_{\rm d}$, the population of H atoms on surfaces is lower due to thermal desorption, which decays the abundance of \ce{CH3NH2}. Thus, \ce{CH3NH2} formation is slightly enhanced on CO ices at low temperatures and $\zeta$. However, the effect of the CO ice in the predicted abundance of \ce{CH3NH2}, is smaller than in the O-bearing COMs described above.

\paragraph{Unaffected Molecules: \ce{CH3OH} (Methanol), \ce{C2H4} (Ethylene), \ce{CH4} (Methane) and \ce{CH3CN} (Methyl cyanide) }

\begin{figure}
    \centering
    \includegraphics[width=\linewidth]{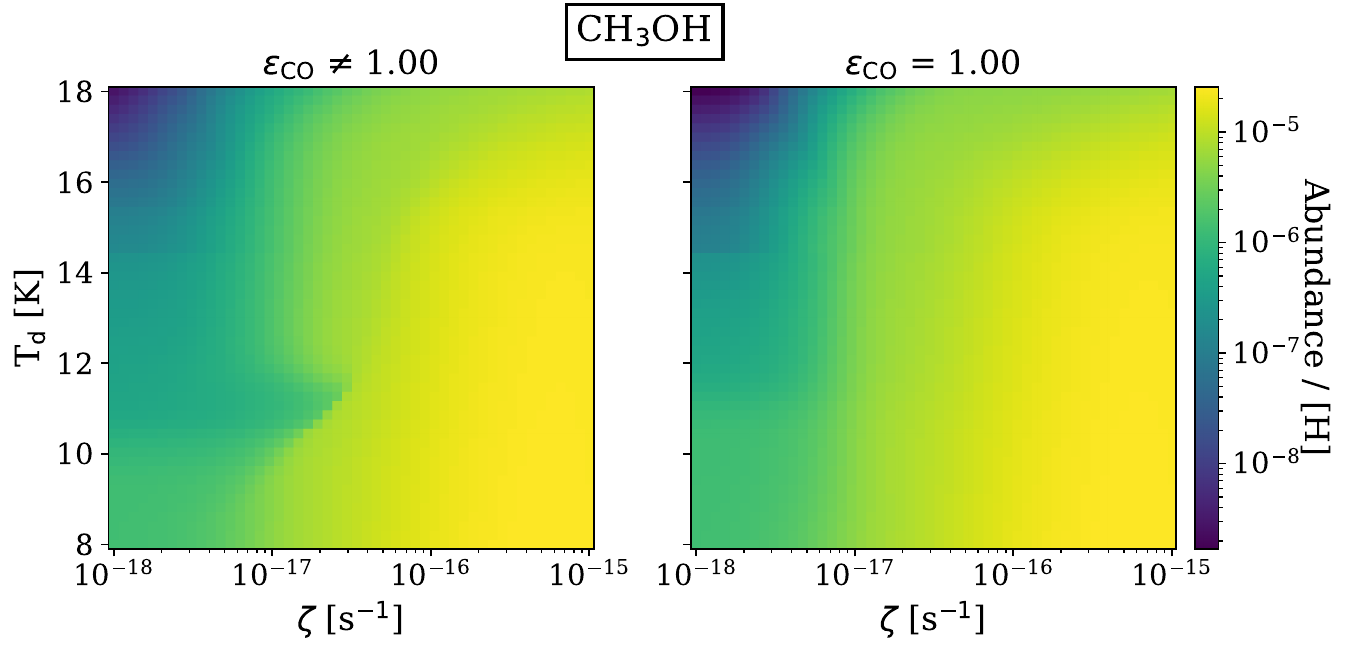} \\
    \includegraphics[width=\linewidth]{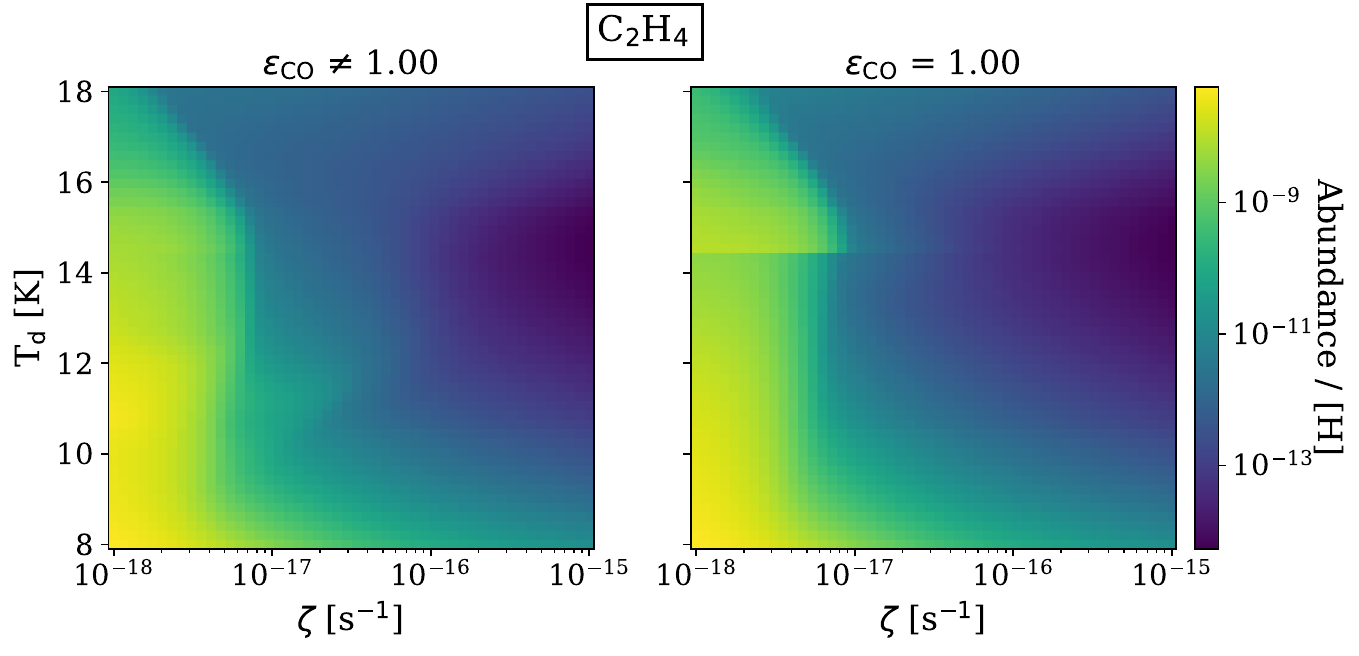} \\
    \includegraphics[width=\linewidth]{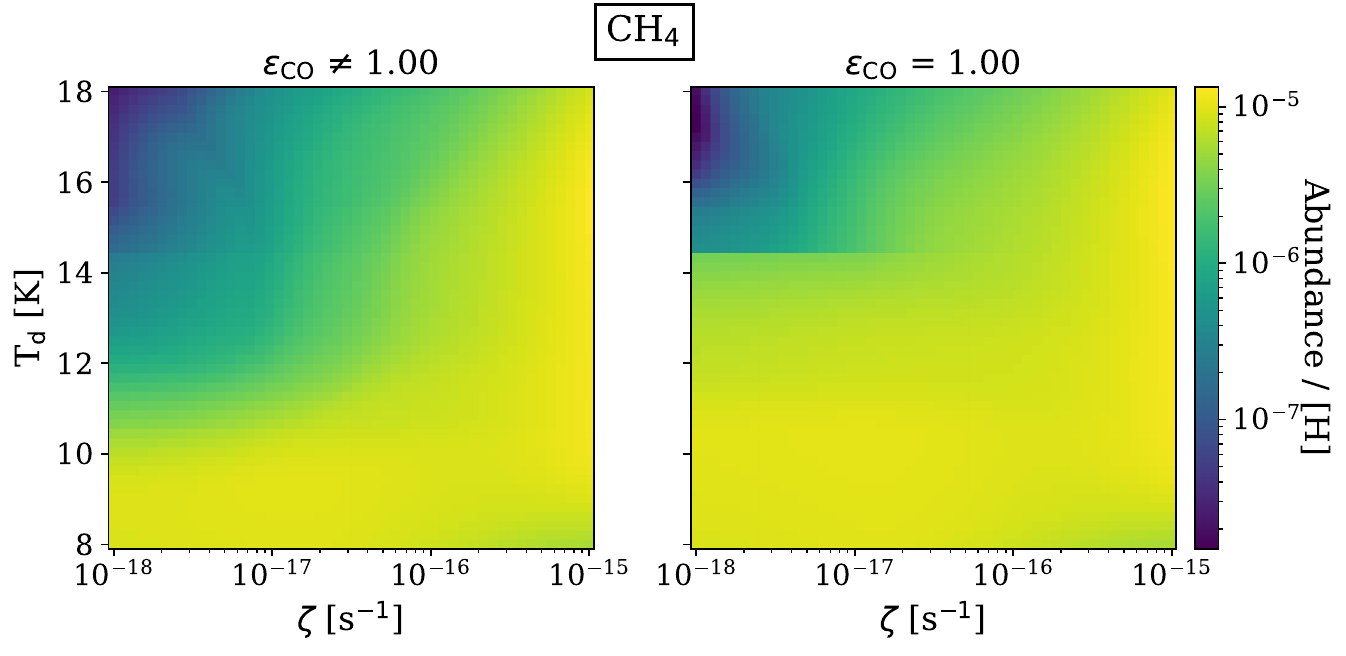} \\
    \includegraphics[width=\linewidth]{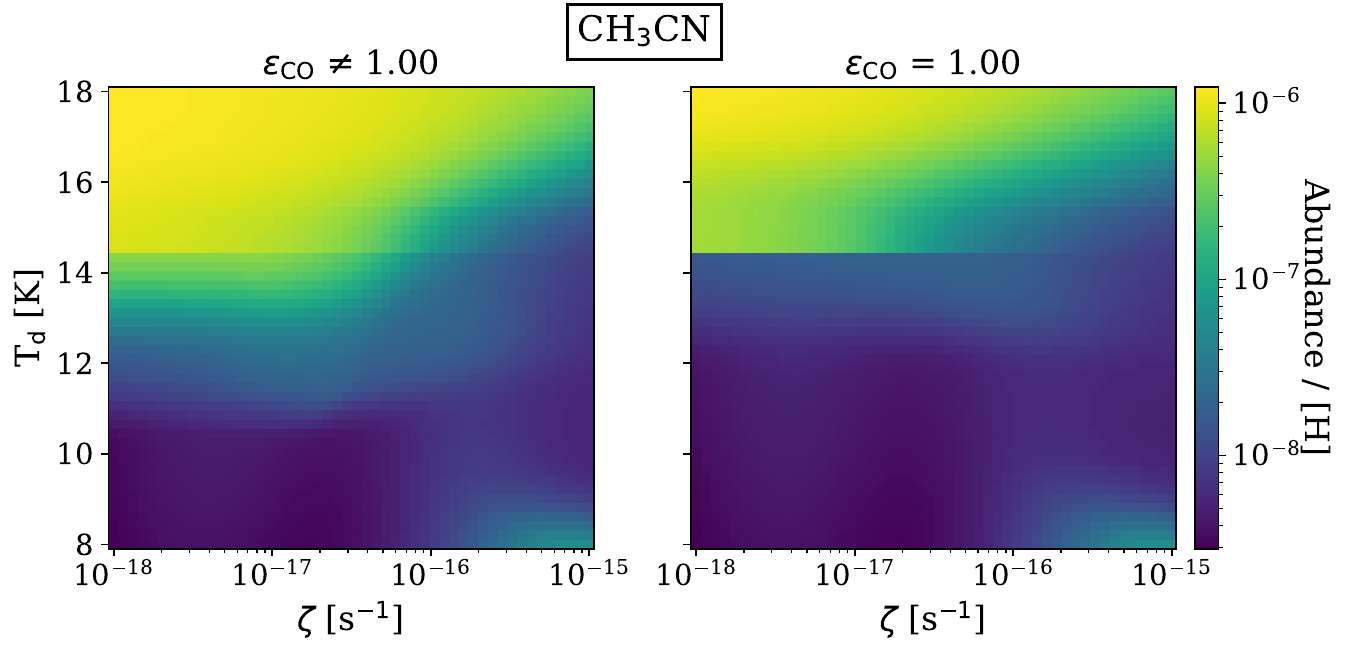} 
    \caption{\textbf{Same as Figure \ref{fig:heatmaps:12K} but for \ce{CH3OH}, \ce{C2H4}, \ce{CH4} and \ce{CH3CN}.}}
    \label{fig:heatmaps:noch3}
\end{figure}

The abundances of  \ce{CH3OH}, \ce{CH3CN}, and \ce{CH4} at time 2$\times$10$^5$ yrs for $\epsilon_{\ce{CO}}$=1.0 and $\epsilon_{\ce{CO}}\neq$1.0 are presented in Figure \ref{fig:heatmaps:noch3}. We observe an almost null effect of the CO ice layer on the abundances, \ce{CH3OH}, \ce{C2H4}, and \ce{CH3CN}, or a decrease in the production of the molecule on CO ice (\ce{CH4}). We found common reasons for all the cases: the competition between hydrogenations and reactions with heavier radicals. For \ce{CH3OH}, the CO hydrogenation (Sequence reactions \ref{eq:CO}) dominates the whole chemistry of the molecule and alternative reactions, like

\begin{chequation}
   \begin{equation}
    \ce{CH3 + OH -> CH3OH}
\end{equation} 
\end{chequation}

\noindent are minor. Similar arguments can be found for \ce{C2H4} and \ce{CH3CN}, whose chemistry is dominated by the chain hydrogenation reactions:

\begin{chequation}
\begin{align}
    \ce{C2H2 &->[+\text{H}] C2H3 ->[+\text{H}] C2H4 ( ->[+\text{H}] C2H5 ->[+\text{H}] C2H6) } \\
    \ce{C2N &->[+\text{H}] HCCN ->[+\text{H}] H2CCN ->[+\text{H}] CH3CN}.
\end{align}
\end{chequation}

\noindent Alternative reactions like

\begin{chequation}
\begin{align}
    \ce{CH2 + CH2 &-> C2H4} \\
    \ce{CH3 + CN &-> CH3CN},
\end{align}
\end{chequation}

\noindent are minor. Interestingly, and as mentioned in the previous section, \ce{C2H6} has a small increase in its abundance on CO ices, in comparison with the related \ce{C2H4} molecule. The reason is that reaction \ref{eq:ch3ch3} contributes to a small extent to the chemistry of \ce{C2H6}, possibly due to the larger abundance of \ce{CH3} radicals on the surface. Finally, the formation of \ce{CH4} is enhanced on \ce{H2O} ice over \ce{CO} ice owing to the lower abundance of \ce{CH3} radicals on the surface that are used in all \ce{CO} enhanced reactions presented above.

The absence of important alternative routes of formation on CO ice for the molecules in this section makes us conclude that enhanced chemistry is only applicable to molecules that cannot be formed by hydrogenation. When hydrogenation is possible, and because the diffusion rate of the H atom on ice does not depend on the ice's surface composition in our models, both the updated model and the classical one yield the same results.

\subsection{Chemistry on CO activated at $\geq$ 14 K} \label{sec:modelHCO}

\begin{figure*}
    \centering
    \includegraphics[width=0.45\linewidth]{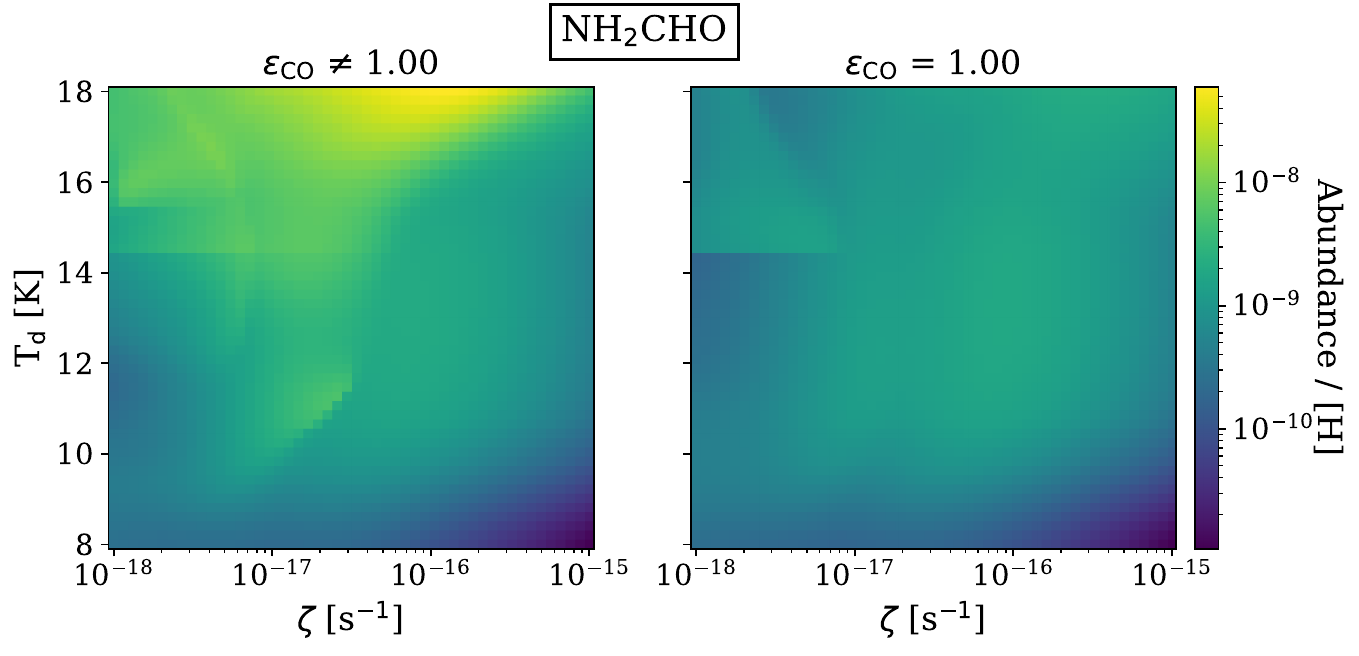}
    \includegraphics[width=0.45\linewidth]{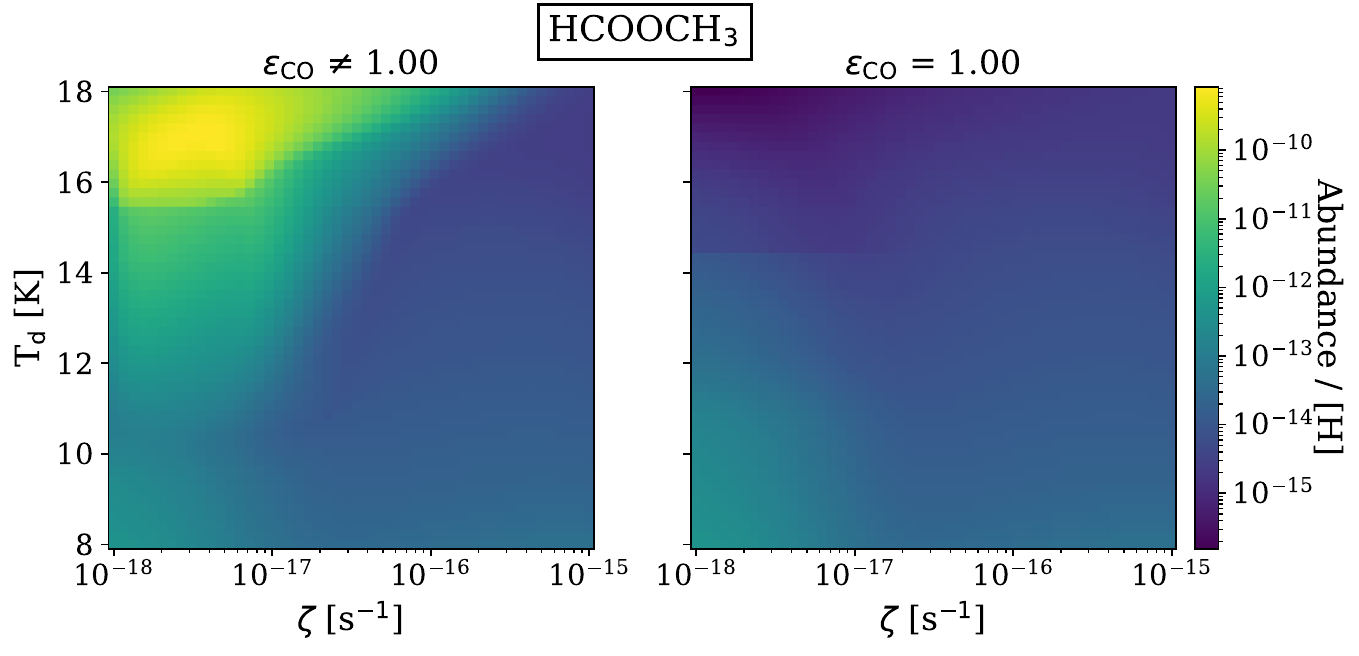} \\
    \includegraphics[width=0.45\linewidth]{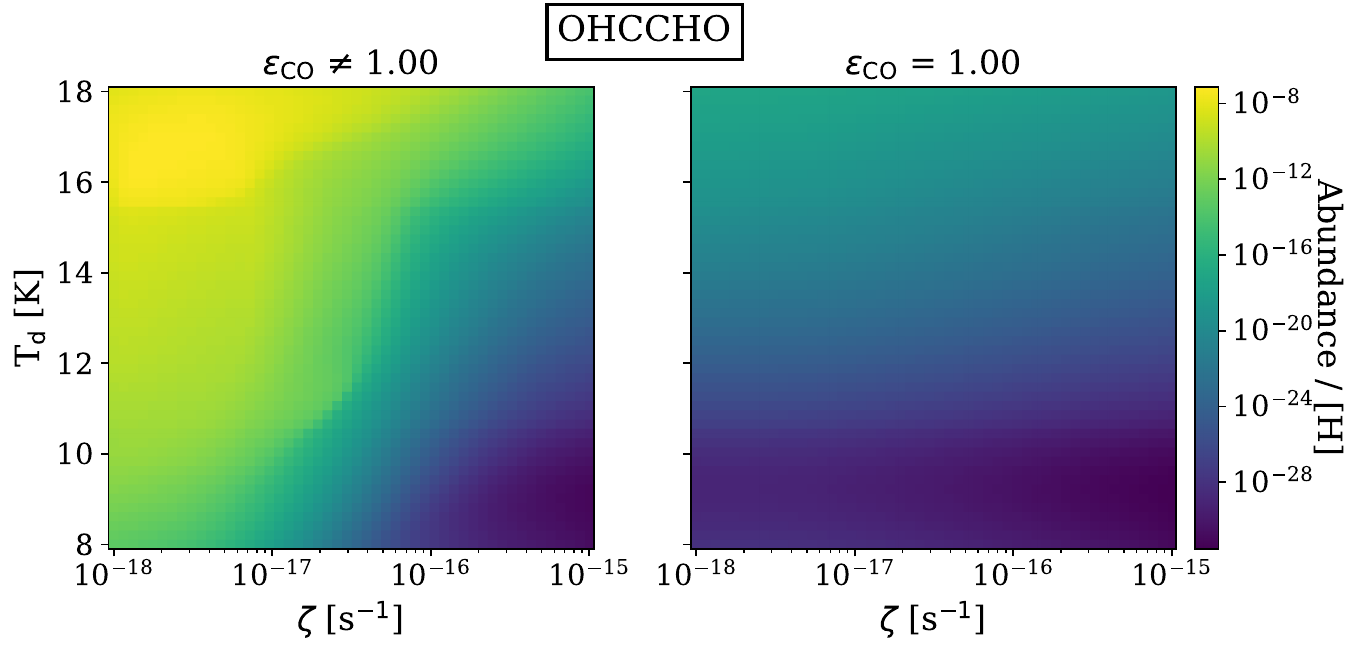} 
    \includegraphics[width=0.45\linewidth]{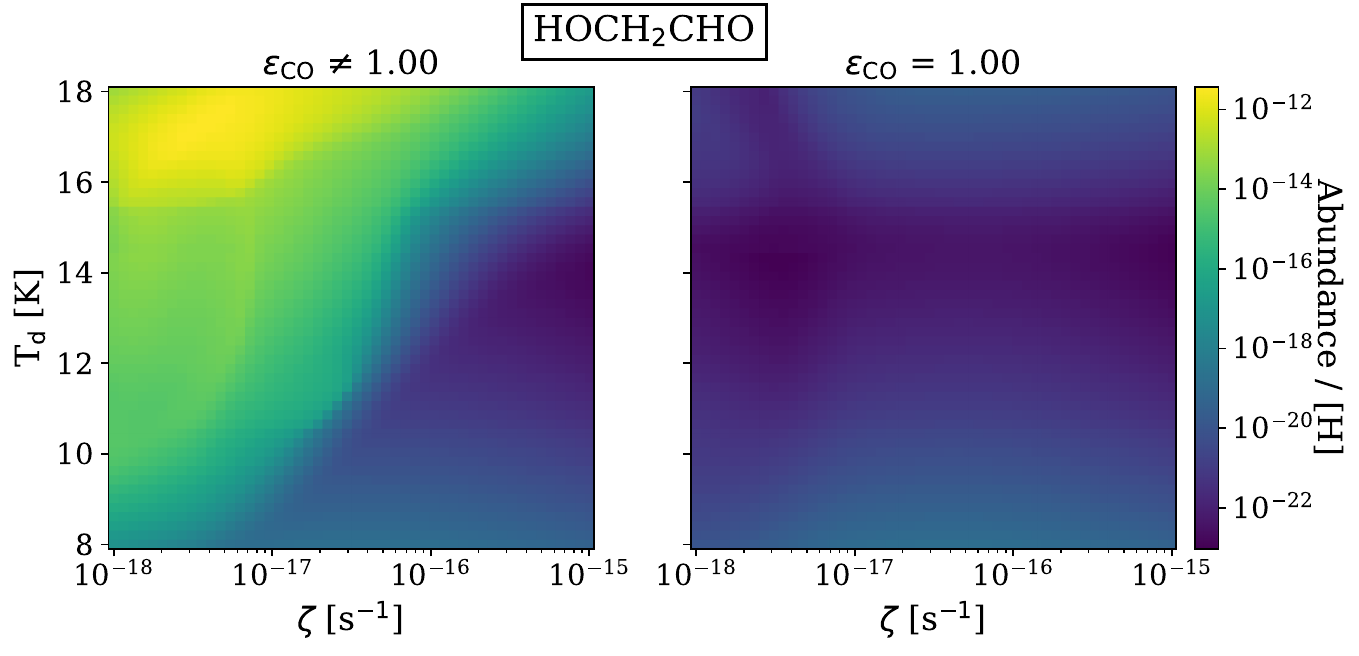} \\
    \includegraphics[width=0.45\linewidth]{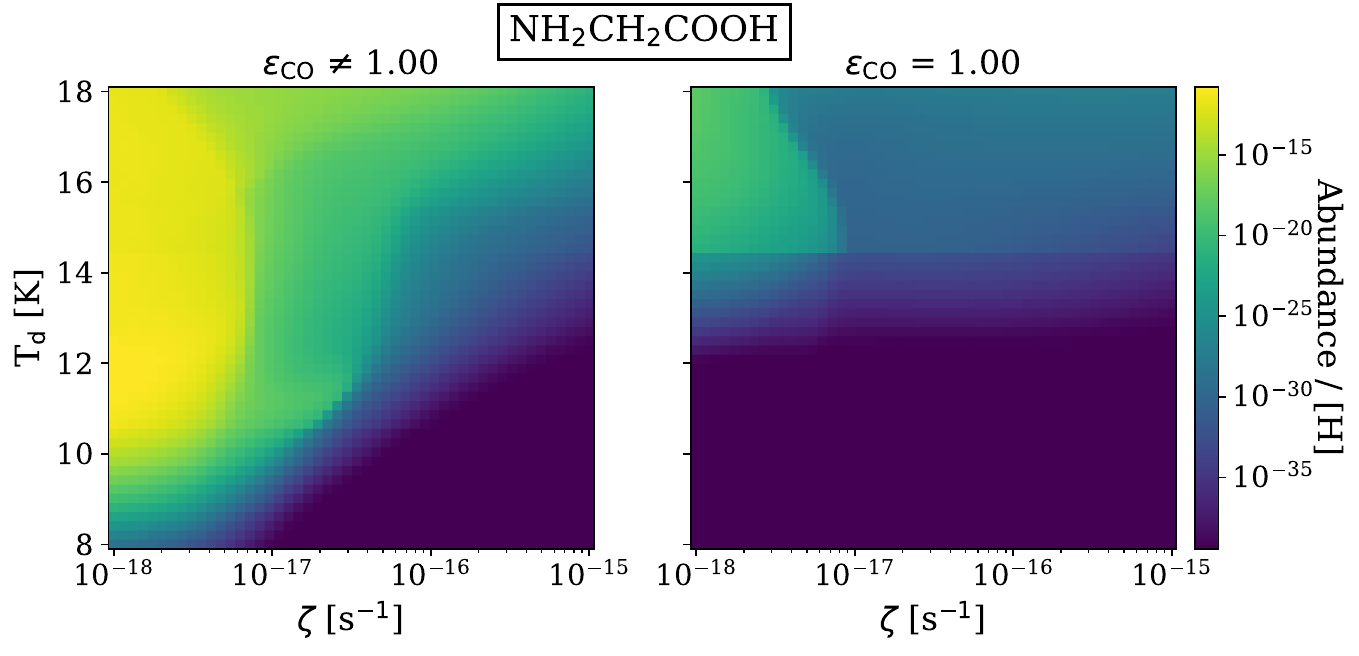}
    \caption{\textbf{Same as Figure \ref{fig:heatmaps:12K} but for \ce{NH2CHO}, \ce{HCOOCH3}, \ce{OHCCHO}, \ce{HOCH2CHO}, and glycine, \ce{NH2CH2COOH}.}}
    \label{fig:heatmaps:14K}
\end{figure*}

The second grand set of molecules affected by the surface composition of the ice substrate refers to molecules related to \ce{HCO} and \ce{NH} that slowly diffuse below 12 K, but start to be rather mobile at $\sim$ 14 K. The impact of CO ice on the chemistry of these molecules is smaller than those presented in Section \ref{sec:modelCOCH3} because \ce{HCO} and \ce{NH} need to compete with the chemistry initiated by \ce{CH3}, \ce{CH2}, \ce{O} and N that have lower diffusion energies and especially by H because both \ce{HCO} and \ce{NH} are easily hydrogenated. The effect of the CO ice is, however, noticeable. We explicitly list some molecules here, namely \ce{NH2CHO} (formamide), \ce{HCOOCH3} (methyl formate), \ce{OHCCHO} (glyoxal), \ce{HOCH2CHO} (glycolaldehyde), and \ce{NH2CH2COOH} (glycine), whose abundances are pictured in Figure \ref{fig:heatmaps:14K}.

The explanation for the enhanced formation on the CO ice is, in this case, common to all molecules except glycine and begins to apply mostly above 12-13 K, coinciding with the start of a slow but noticeable diffusion of HCO and NH (Figure \ref{fig:hopping}). This makes an itemized description of the reactivity, as in the previous section, reiterative. The enhanced abundance of these molecules arises from radical couplings between the parent radicals and HCO/NH. For the species portrayed in Figure \ref{fig:heatmaps:14K} these reactions are, where the competing reactions are in parentheses:

\begin{chequation}
\begin{align}
    \ce{NH2 + HCO &-> NH2CHO} \\
    (\ce{NH2 + HCO &-> NH3 + CO}) \\
    \ce{HCO + CH3O &-> HCOOCH3} \\
    (\ce{HCO + CH3O &-> CH3OH + CO}) \\
    \ce{HCO + HCO &-> OHCCHO} \\
    (\ce{HCO + HCO &-> H2CO + CO}) \\
    \ce{HCO + CH2OH &-> HOCH2CHO} \\
    (\ce{HCO + CH2OH &-> 2H2CO}) 
\end{align}
\end{chequation}

\noindent The dependence on $T_{\rm d}$ is now clearer to explain; all chemistry is activated by the diffusion of \ce{HCO}/\ce{NH2}. On the other hand, the $\zeta$ dependence is precisely the same as in Section \ref{sec:modelCOCH3}. With more available H atoms, diffusive chemistry on CO ices converges to diffusive chemistry on \ce{H2O},\footnote{This generalization holds when assuming similar activation energies for reactions on both \ce{H2O} and CO surfaces. The limits of the applicability of this approximation will be explored in further works.} because heavy radical diffusion is hampered by the hydrogenation of the radicals. The small differences between the molecules in Figure \ref{fig:heatmaps:14K} are attributed to subtle differences in the abundances of parent molecules and precursors at the given $T_{\rm d}$. It is more difficult to pinpoint the exact reason for each case. 

Even with the increased formation of COMs on CO ice, the species whose parent species is HCO have lower abundances than the ones related with \ce{CH3}/\ce{CH2}. This is especially true for \ce{HCOOCH3} and \ce{HOCH2CHO}. For \ce{NH2CHO}, an important prebiotic precursor \citep{lopez-sepulcre_interstellar_2019}, the abundances predicted by the CO-rich model bring the ice abundance of \ce{NH2CHO} to values up to $\sim$ 5$\times$10$^{-9}$ at 14 K and canonical $\zeta$ (1.3$\times$10$^{-17}$ s$^{-1}$). We found a narrow maximum at 18 K and $\zeta$=1$\times$10$^{-16}$ s$^{-1}$ with fractional abundances of 5$\times$10$^{-8}$, which indicate the ideal conditions for the production of this molecule. However, at the maximum, the $T_{\rm d}$ is rather close to the CO desorption temperature \citep{Bisschop2006, Fuchs2006}. The high reference binding energy of \ce{NH2CHO}, 5400 K in our model and in reality likely even higher, challenges the return of the molecule to the gas phase in prestellar cores. Therefore, we encourage non-thermal desorption studies of \ce{NH2CHO} on CO ice to confirm the viability of releasing this molecule to the gas phase or search with JWST observations.

In the case of glycine, one of the interstellar molecules without positive detection that attracts more attention as the simplest amino acid, it has a low-temperature synthetic route on CO:

\begin{chequation}
\begin{align}
    \ce{HOCO + CH2 &-> H2CCOOH} \\
    \ce{H2CCOOH + NH &-> NHCH2COOH} \\
    \ce{NHCH2COOH + H &-> NH2CH2COOH}
\end{align}
\end{chequation}

\noindent e.g. a three-step synthesis. In this case, the formation of the molecule is allowed through the concomitant diffusion of the reactants and the formation of HOCO (\ref{eq:hoco}). Despite an increase in the formation of glycine by many orders of magnitude, the absolute abundance of glycine in ice remains too small for potential detection. However, we refer to \cite{Ioppolo2020} for alternative routes relying on non-thermal schemes that could potentially bring the abundances closer to detectable amounts.

\textbf{Finally, we summarize and merge all the results discussed for ice abundances in Table \ref{tab:summaryRevision} for a semiquantitative estimation of the effect of CO ice in the chemistry of the molecules considered in these Sections. The table reinforces our general conclusion that COM abundances are enhanced in the models of $\epsilon_{\ce{CO}}$$\neq$1.0 compared with those of $\epsilon_{\ce{CO}}$=1.0 in most of the combinations of $\zeta$ and $T_{\rm d}$. The enhancement, however, diminishes in the case of high $\zeta$ and low $T_{\rm d}$.}

\begin{table*}[t]
\begin{center}
\caption{\textbf{Ice abundances at t=2$\times$10$^5$ yrs with respect to hydrogen nuclei considering Equation \ref{eq:scale} (i.e. $\epsilon_{\ce{CO}}$ $\neq$1.0) or without considering it. The different levels for $\zeta$ and $T_{\rm d}$ are $\mathcal{L}$ (low), $\mathcal{M}$ (medium), and $\mathcal{H}$ (high). They correspond to $\mathcal{L}$ $\zeta$=1.0$\times$10$^{-18}$ s$^{-1}$, $\mathcal{M}$ $\zeta$=1.3$\times$10$^{-17}$ s$^{-1}$ (as the nominal $\zeta$), $\mathcal{H}$ $\zeta$=5.0$\times$10$^{-16}$ s$^{-1}$, $\mathcal{L}$ $T_{\rm d}$=8 K, $\mathcal{M}$ $T_{\rm d}$=13 K and $\mathcal{H}$ $T_{\rm d}$=18 K. The A(B) notation is used to denote A$\times$10$^{B}$.   }}
\label{tab:summaryRevision}
\resizebox{\textwidth}{!}{%
\begin{tabular}{l|ccccccccc}
\toprule
Species & $\mathcal{L}$ $\zeta$, $\mathcal{L}$ $T_{\rm d}$ & $\mathcal{M}$ $\zeta$, $\mathcal{L}$ $T_{\rm d}$ & $\mathcal{H}$ $\zeta$, $\mathcal{L}$ $T_{\rm d}$ & $\mathcal{L}$ $\zeta$, $\mathcal{M}$ $T_{\rm d}$ & $\mathcal{L}$ $\zeta$, $\mathcal{H}$ $T_{\rm d}$ & $\mathcal{M}$ $\zeta$, $\mathcal{M}$ $T_{\rm d}$ &  $\mathcal{M}$ $\zeta$, $\mathcal{H}$ $T_{\rm d}$ & $\mathcal{H}$ $\zeta$, $\mathcal{M}$ $T_{\rm d}$ & $\mathcal{H}$ $\zeta$, $\mathcal{H}$ $T_{\rm d}$ \\
\bottomrule
\ce{CH3CHO} & 9.8(-11) / 2.9(-12) & 4.3(-12) / 3.7(-12) & 5.9(-12) / 6.0(-12) & 6.1(-9) / 2.1(-12) & 3.6(-9) / 1.2(-10) & 1.6(-8) / 4.2(-12) & 1.6(-8) / 3.4(-11) & 3.0(-12) / 2.8(-12) & 1.4(-8) / 2.4(-11) \\
\ce{CH3OCH3} & 8.1(-13) / 5.0(-14) & 3.5(-14) / 3.5(-14) & 7.9(-14) / 7.9(-14) & 2.3(-11) / 7.8(-15) & 6.6(-11) / 2.7(-12) & 1.6(-9) / 1.3(-14) & 2.7(-9) / 2.6(-14) & 6.3(-14) / 6.2(-14) & 3.0(-11) / 2.1(-14) \\
\ce{C2H5OH} & 4.3(-14) / 1.4(-15) & 1.4(-14) / 1.2(-14) & 2.8(-14) / 2.8(-14) & 1.7(-12) / 2.3(-16) & 2.7(-13) / 1.3(-14) & 5.8(-11) / 2.3(-15) & 6.1(-11) / 1.5(-15) & 5.4(-15) / 5.3(-15) & 1.5(-12) / 6.2(-15) \\
\ce{CH3COOH} & 1.7(-20) / 1.1(-35) & 1.2(-25) / 1.2(-34) & 2.4(-33) / 1.1(-33) & 2.6(-10) / 3.4(-19) & 6.7(-10) / 2.3(-10) & 5.0(-10) / 1.1(-21) & 3.5(-9) / 6.6(-13) & 4.7(-20) / 3.9(-23) & 9.4(-10) / 1.4(-12) \\
\ce{C2H6} & 5.0(-7) / 4.9(-7) & 4.7(-7) / 4.7(-7) & 8.8(-7) / 8.8(-7) & 7.8(-7) / 2.8(-7) & 9.3(-7) / 1.1(-6) & 1.3(-6) / 3.6(-7) & 1.0(-6) / 1.3(-6) & 9.0(-7) / 8.9(-7) & 1.5(-6) / 1.3(-6) \\
\ce{CH3NH2} & 9.4(-10) / 6.6(-10) & 1.3(-9) / 1.2(-9) & 2.3(-9) / 2.3(-9) & 8.8(-7) / 3.1(-6) & 9.3(-9) / 8.2(-8) & 1.0(-6) / 2.3(-6) & 7.8(-8) / 2.8(-7) & 4.5(-7) / 2.8(-7) & 4.9(-7) / 5.5(-7) \\ 
\ce{CH3OH} & 1.4(-6) / 1.4(-6) & 6.1(-6) / 6.0(-6) & 2.5(-5) / 2.5(-5) & 3.5(-7) / 4.1(-7) & 2.8(-9) / 1.7(-9) & 2.6(-6) / 4.9(-6) & 8.2(-7) / 1.3(-6) & 2.4(-5) / 2.4(-5) & 7.3(-6) / 6.1(-6) \\ 
\ce{C2H4} & 5.4(-8) / 5.7(-8) & 1.8(-9) / 1.9(-9) & 2.4(-11) / 2.4(-11) & 1.4(-8) / 7.4(-9) & 8.0(-11) / 3.2(-10) & 6.9(-12) / 1.2(-12) & 1.8(-12) / 3.3(-12) & 2.0(-14) / 2.0(-14) & 3.7(-13) / 4.8(-13) \\ 
\ce{CH4} & 9.1(-6) / 9.1(-6) & 9.6(-6) / 9.6(-6) & 6.0(-6) / 6.0(-6) & 4.9(-7) / 5.9(-6) & 2.6(-8) / 1.9(-8) & 1.6(-6) / 6.9(-6) & 5.4(-7) / 1.0(-6) & 9.3(-6) / 1.0(-5) & 5.2(-6) / 6.6(-6) \\
\ce{CH3CN} & 3.0(-9) / 2.9(-9) & 3.3(-9) / 3.2(-9) & 4.5(-8) / 4.5(-8) & 4.7(-8) / 7.2(-9) & 1.2(-6) / 1.2(-6) & 5.8(-8) / 8.5(-9) & 1.1(-6) / 9.7(-7) & 1.0(-8) / 7.9(-9) & 5.0(-7) / 3.5(-7) \\
\ce{NH2CHO} & 2.6(-10) / 2.4(-10) & 2.0(-10) / 1.8(-10) & 2.3(-11) / 2.3(-11) & 4.8(-10) / 2.3(-10) & 4.6(-9) / 4.9(-10) & 2.8(-9) / 1.3(-9) & 1.9(-8) / 9.1(-10) & 9.2(-10) / 9.1(-10) & 1.5(-8) / 2.1(-9) \\
\ce{HCOOCH3} & 4.2(-13) / 4.2(-13) & 3.9(-14) / 3.7(-14) & 4.5(-14) / 4.5(-14) & 5.8(-13) / 2.7(-14) & 3.2(-11) / 1.5(-16) & 1.2(-13) / 6.0(-15) & 1.4(-10) / 5.6(-16) & 8.7(-15) / 8.5(-15) & 6.9(-15) / 1.9(-15) \\
\ce{OHCCHO} & 4.9(-14) / 7.8(-29) & 1.7(-19) / 3.0(-29) & 1.1(-30) / 3.6(-31) & 3.2(-10) / 1.4(-23) & 6.4(-9) / 6.3(-18) & 5.6(-12) / 4.4(-24) & 5.9(-9) / 3.2(-18) & 2.6(-24) / 7.5(-26) & 1.1(-13) / 3.8(-19) \\
\ce{HOCH2CHO} & 6.5(-18) / 1.0(-20) & 1.3(-19) / 1.1(-19) & 6.0(-20) / 5.8(-20) & 1.2(-14) / 6.7(-23) & 8.1(-14) / 8.6(-22) & 1.6(-15) / 1.0(-22) & 1.4(-12) / 1.2(-20) & 5.4(-23) / 5.3(-23) & 2.1(-16) / 1.3(-20) \\
\ce{NH2CH2COOH} & 2.7(-29) / 3.8(-40) & 3.9(-40) / 3.8(-40) & 3.7(-40) / 3.7(-40) & 4.9(-12) / 5.1(-34) & 1.5(-12) / 7.9(-19) & 5.3(-20) / 1.5(-38) & 3.1(-16) / 1.5(-28) & 1.8(-36) / 1.6(-39) & 2.2(-20) / 3.9(-28) \\ 
\bottomrule
\end{tabular}
}
\end{center}
\end{table*}

\subsection{Impact on gas phase abundances} \label{sec:gas}

We briefly discuss how CO ice affects the desorption of some molecules into the gas phase. Lower binding energies on CO-dominated ice surfaces can potentially enhance a series of thermal and nonthermal desorption mechanisms. From the three nonthermal mechanisms considered in our models, photodesorption is the only formally independent of the adsorbates' binding energy, at least in our models. The remaining two, cosmic-ray-induced and chemical desorption, explicitly depend on the binding energy. The former is simulated in the Hasegawa-Herbst formulation \citep{hasegawa_three-phase_1993}, which means that every grain is briefly heated stochastically to 70 K (10 $\mu$s) and the update in BE affects the desorption rates of molecules. With respect to chemical desorption, in Section \ref{sec:methods} we discuss that our models use the same chemical desorption probability on \ce{H2O} ice and \ce{CO} ice because the literature does not have experimental or theoretical data on the partition of the reaction energy into CO ices. Furthermore, the chemical desorption fraction ($f$, equation \ref{eq:cd_frac}) will also depend on the molecule under consideration \citep{Fredon2021}. Therefore, there are severe uncertainties in the probability of chemical desorption, and we prefer not to enlarge them in our models. However, we note that if $f$ is calculated as a function of CO surface coverage, the differences in gaseous molecular abundances between models with $\epsilon_{\ce{CO}}$=1.0 and $\epsilon_{\ce{CO}}\neq$1.0 should be larger.

\begin{figure*}
    \centering
    \includegraphics[width=0.45\linewidth]{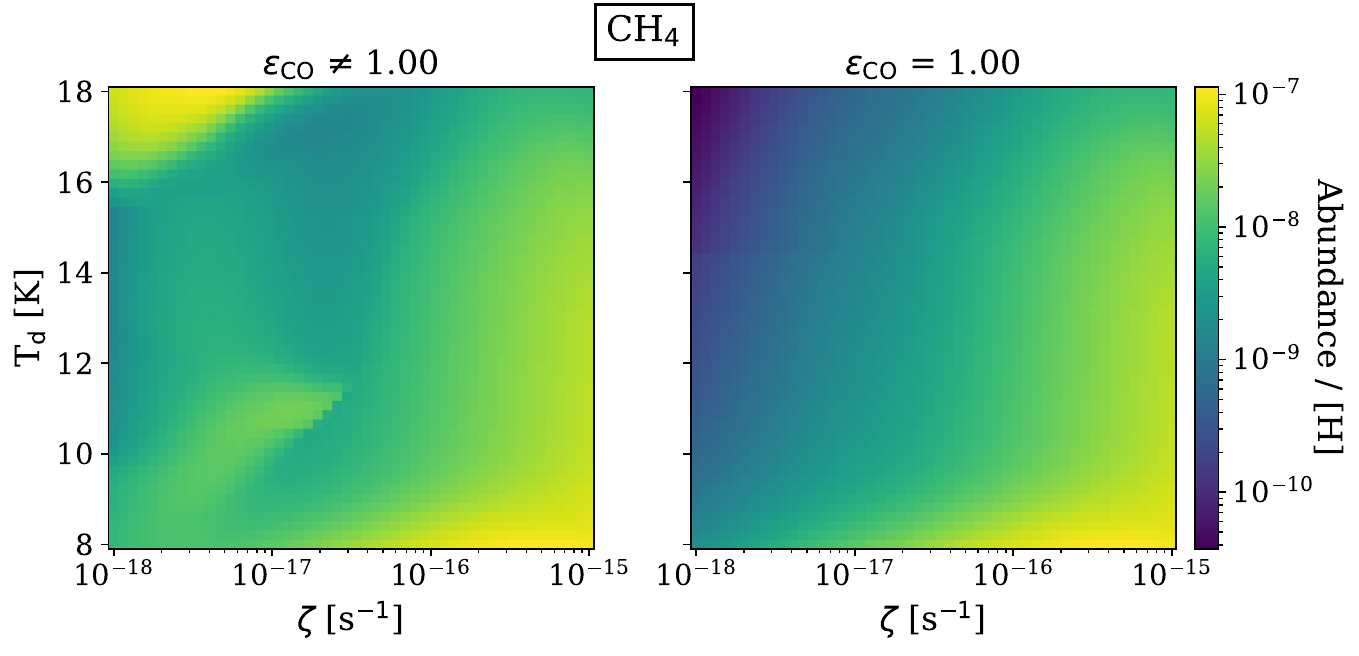} 
    \includegraphics[width=0.45\linewidth]{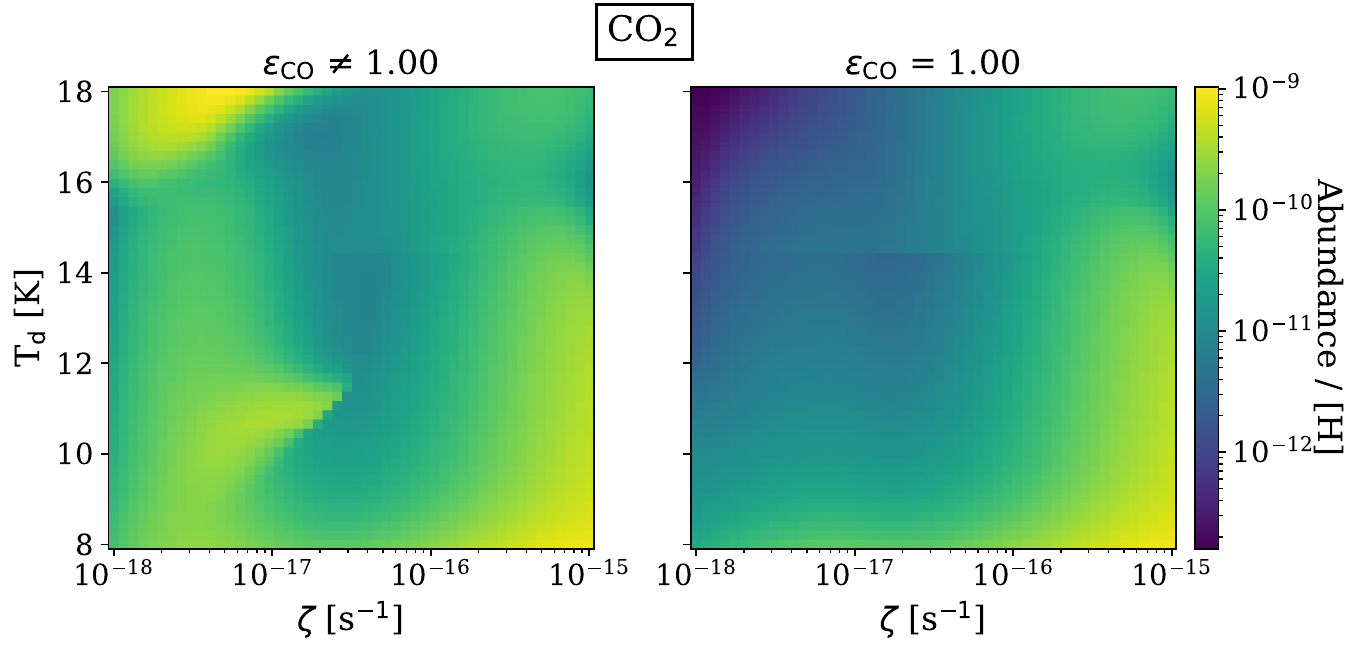} \\
    \includegraphics[width=0.45\linewidth]{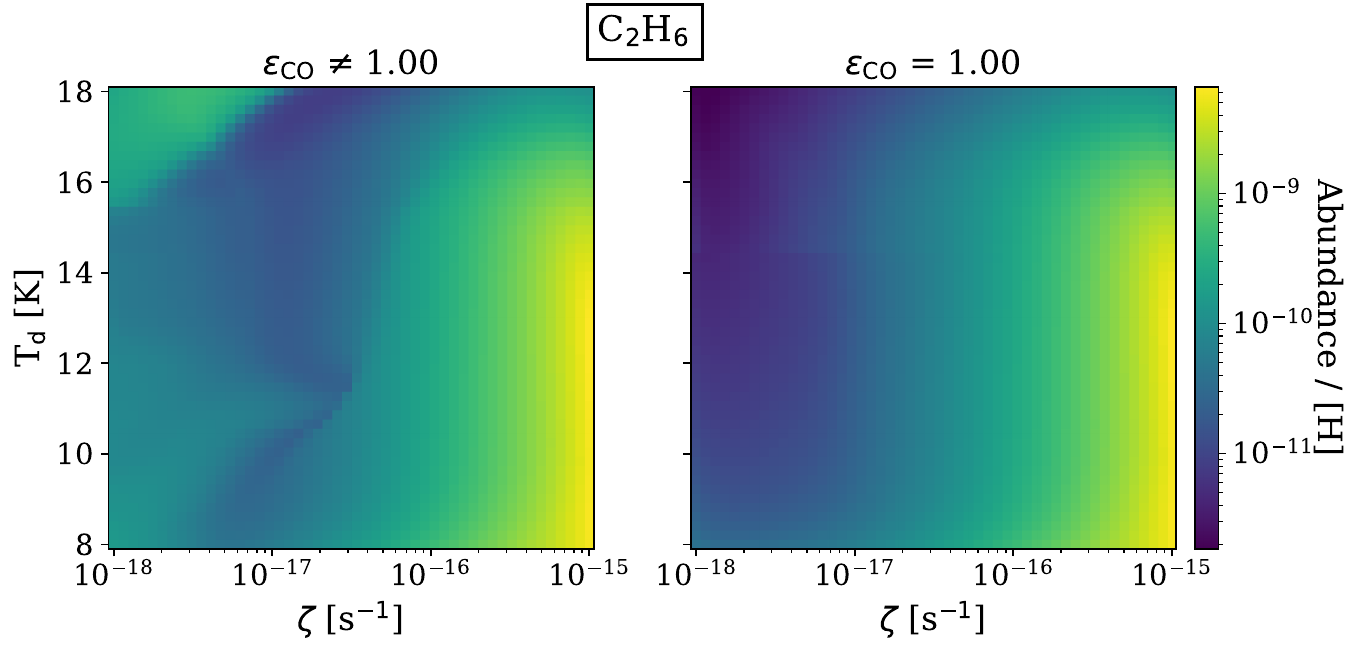}
    \includegraphics[width=0.45\linewidth]{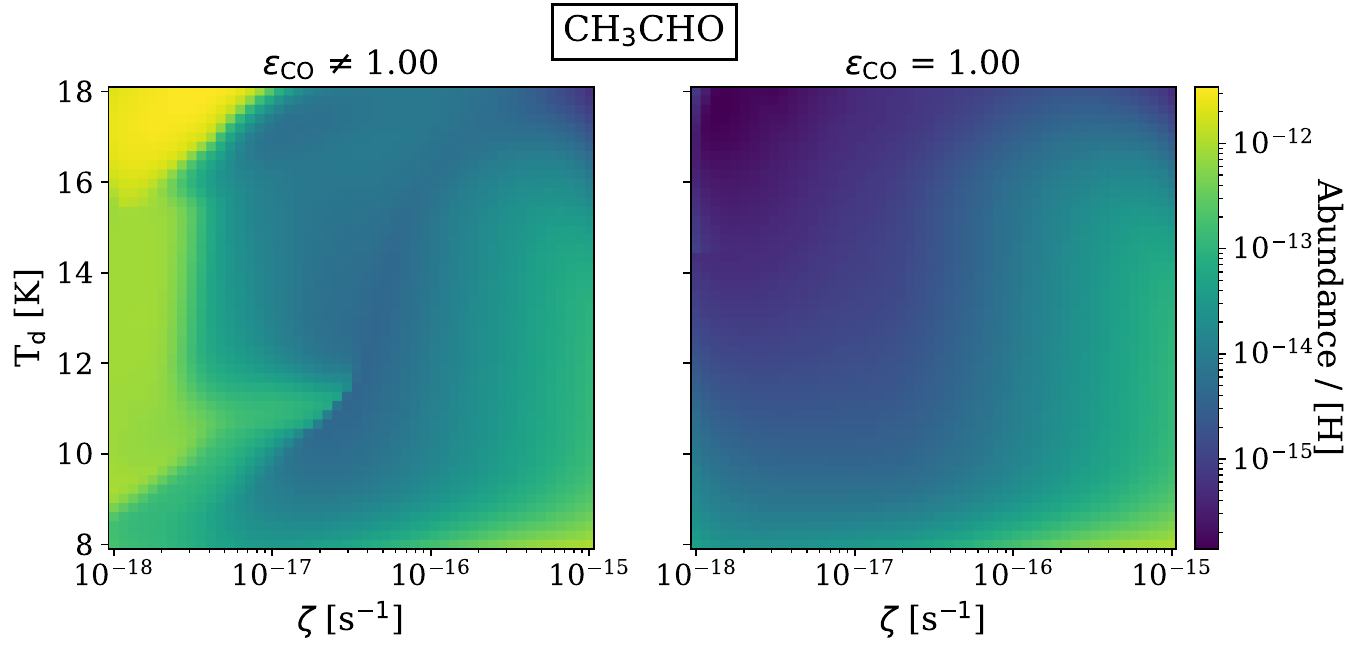} \\
    \includegraphics[width=0.45\linewidth]{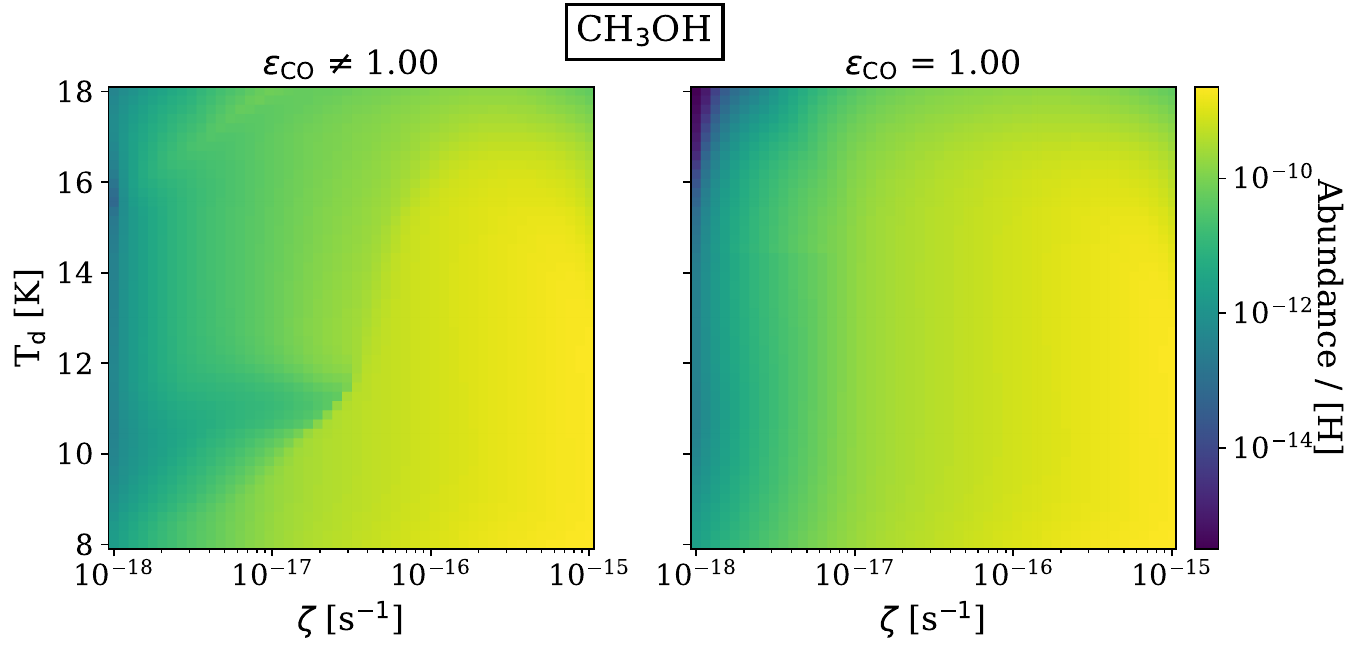}
    \caption{\textbf{Gas abundances at t=2$\times$10$^5$ yrs (see text for an explanation of chemical timescales for gaseous molecules) in models of various $\zeta$ and $T_{\rm d}$. The selected molecules are \ce{CH4}, \ce{CO2}, \ce{C2H6}, \ce{CH3CHO}, and \ce{CH3OH}.}}
    \label{fig:heatmaps:gas}
\end{figure*}

We study the effect on gas phase abundances of two small molecules (\ce{CH4} and \ce{CO2}) and three COMs (\ce{C2H6}, \ce{CH3CHO} and \ce{CH3OH}). Figure \ref{fig:heatmaps:gas} shows the gas-phase abundances of these molecules at $t$=2$\times$ 10$^{5}$ years as a function of $\zeta$ and $T_{\rm d}$. In contrast to ices at low temperatures, where molecules are stored and protected in the mantle, the gas is in a harsher environment where molecules do not reach a steady state, and are continuously formed and destroyed. Therefore, there is no unique choice of time to collect our data, and $t$=2$\times$10$^{5}$ yr does not reflect the ideal time for detecting the molecule, and the abundances presented here should not be considered as predictions, but rather as a comparison to show the differences in the models when desorption and diffusion rates vary with the composition of the ice surface. \textbf{This is the reason, combined with the complexities in the gas-phase reaction network, why we refrain from providing a Table similar to Table \ref{tab:summaryRevision} for gas-phase species. }

The increased abundance of gas-phase molecules is hardly related to a single physical or chemical process; for example, gas-phase abundances are much more sensitive to destruction by gas-phase reactions as mentioned above. We first observe a smaller dust temperature dependence of gaseous molecular abundances to ice abundances, except for CO at high $T_{\rm d}$. This is indirect evidence for the dominance of nonthermal effects in desorption at 8--18 K. For example, with low $\zeta$, the gaseous \ce{CH4} abundance at $T_{\rm d}$ $<$ 18 K is higher in the model with $\epsilon_{\ce{CO}}\neq$1.0 due to the lower BE, which improves CR-induced desorption. However, grain contributions compete with the following gas-phase destruction reaction:

\begin{chequation}
\begin{equation}
    \ce{CH4 + H3+ -> CH5+ + H2},   
\end{equation}
\end{chequation}

\noindent at intermediate $\zeta$. At high $\zeta$, the higher abundance of \ce{CH4} desorbed from ice (see Section \ref{sec:modelCOCH3}) dominates over the above reaction. This is an example of the complex reasons behind the gas-phase molecular abundances. Similar arguments can be given for \ce{CO2}, where competing reactions are at play in the predicted abundances. The increased abundance of the gas phase \ce{CO2} in models at low $\zeta$ and $\epsilon_{\ce{CO}}\neq$1.0 is not due to the desorption of \ce{CO2} from the grain, but rather to the increased formation in the gas phase through:

\begin{chequation}
\begin{align}
    \ce{H3+ + CO &-> HCO+ + H2} \\
    \ce{HCO+ + OH &-> HCO2+ + H} \\
    \ce{HCO2+ + CO &-> HCO+ + CO2} \\
    \ce{HCO2+ + e- &-> H + CO2} \\
    \ce{HCO2+ + NH3 &-> NH4+ + CO2}
\end{align}
\end{chequation}

\noindent so it depends indirectly on the amount of CO in the gas, which increases thanks to cosmic ray induced desorption at low and intermediate $\zeta$ (e.g. when hydrogenation of CO ice is slower), as shown above for \ce{CH4}. This highlights once again the complex and hardly predictable grain-gas interaction as well as the importance of chemical models in addressing this complex chemistry.

For most of the COMs, the main non-thermal desorption mechanism is chemical desorption, as explained above. For \ce{C2H6}, cosmic ray-induced desorption and chemical desorption are on par, owing to the low binding energy of \ce{C2H6}. In the case of \ce{CH3CHO}, whose ice abundance is most affected by the binding energy update on the CO ice (see Section \ref{sec:modelCOCH3}), the explanation for the enhanced gas-phase abundance with $\epsilon_{\ce{CO}}\neq$1.0 models is straightforward. Chemical desorption is efficient because \ce{CH3CHO} has relatively low binding energy \citep{molpeceres_desorption_2022, ferrero_acetaldehyde_2022}, well represented by the \cite{Garrod2013} reaction network (2500 K). The increased abundance on the ices presented in Section \ref{sec:modelCOCH3} directly correlates with the observed abundance in the gas, considering that our chemical desorption scheme does not change with $\epsilon_{\ce{CO}}$. Gas abundances of molecules with intermediate binding energies, for example, \ce{CH3OCH3}, are also enhanced. However, we note that the abundances of \ce{CH3CHO} are low for its detection, but that is a consequence of the choice of time for the extraction of the data, as mentioned above. As an example, the global maximum in gas \ce{CH3CHO} abundances in our models corresponds to the model with $\epsilon_{\ce{CO}}\neq$1.00, $T_{\rm d}$=10.7 K and $\zeta$=1$\times$10$^{-18}$ s$^{-1}$, at t=3.2$\times$10$^{4}$ yr, with a fractional abundance of 1.2$\times$10$^{-10}$ with respect to H, e.g. a detectable amount consistent with prestellar core observations for this molecule \citep{JImenezSerra2016, megias_complex_2022}.

Finally, it is worth indicating that our chemical models for $\epsilon_{\ce{CO}}\neq$1.00 do not show a notable increase in the abundance of \ce{CH3OH} in the gas phase (Figure \ref{fig:heatmaps:gas}, bottom panel), as expected from the similar abundances of ice on the \ce{H2O} and \ce{CO} ices reported in Section \ref{sec:modelCOCH3}. An in-depth study of chemical desorption of \ce{CH3OH} from CO ices is desirable, to improve our description of chemical desorption and complement other desorption mechanisms available in the literature, e.g., photodesorption, co-desorption, and recently resonant infrared photodesorption; \cite{Oberg2009, Bertin2016,MartinDomenech2016, ligterink_methanol_2018, Santos2023}.

\section{Discussion and Conclusions} \label{sec:closing}

Using state-of-the-art astrochemical models, we have investigated the role of reduced binding energies of adsorbates on CO-dominated ices in diffusion-mediated astrochemical reaction networks. The enhanced chemistry we found is attributed solely to the change in the intermolecular forces that drive diffusion, and new reactivity is not explicitly addressed (see, for example, the exclusive CO ice Eley-Rideal reaction, \ce{C + CO -> CCO}; \cite{Fedoseev2022, ferrero2023formation}). Diffusive chemistry is affected by CO ices, although non-diffusive mechanisms can also play a significant role in the promotion of chemistry in interstellar grains \citep{Jin2020, Garrod2022}. In fact,our models share some trends with the non-diffusive scenario, where \citet{Jin2020} report efficient production of \ce{CH3CHO} and \ce{CH3OCH3}. Although our proposed routes differ, our findings are complementary for \ce{CH3CHO} and \ce{CH3OCH3}. The non-diffusive mechanism relies upon the reactions \ce{CH3$^{*}$ + CO} and \ce{CH3$^{*}$ + H2CO} (where the asterisk denotes a translational excited radical) for reactions to form \ce{CH3CO} and \ce{CH3OCH2} that can be easily hydrogenated to \ce{CH3CHO} and \ce{CH3OCH3}.

Our chemical models reveal that the importance of CO-dominated ice is highly dependent on the physical conditions of the core, mainly through the dust temperature $T_{\rm d}$ and the cosmic ray ionization rate $\zeta$. From these two, the critical parameter for enhanced chemistry on CO ices is $\zeta$, which can vary several orders of magnitude \citep{dalgarno_galactic_2006, padovani_cosmic-ray_2009, Padovani2018, padovani_production_2018}. For example, in the well studied prestellar core L1527, the cosmic ray ionization rate is likely to be closer to the canonical value of $\zeta$=1.3$\times$10$^{-17}$ s$^{-1}$ \citep{redaelli_cosmic-ray_2021} while in the giant molecular cloud G+0.693-0.027 in the center of our galaxy, $\zeta$ may present values between 100 and 1000 higher than the canonical value \citep{goto_h_2013, Goto2013}. In general, COM formation is enhanced on CO-dominated ice with low $\zeta$, while chemistry on CO-dominated ices and \ce{H2O}-dominated ices converges at very high $\zeta$ and \textbf{low $T_{\rm d}$}. However, the effect at intermediate values of $\zeta$ is not trivial to predict, and astrochemical models remain instrumental in unraveling each specific case. 

Our work comes at a good time for the astrochemistry canvas, with JWST observations indicating an important component of COM before forming a protostar \citep{McClure2023}. However, the definitive identification of COMs other than \ce{CH3OH} in ice observations has proven challenging \cite{Nazari2024}. Our results encourage us to focus on the search for COMs formed from the combination of CO-derived radicals \ce{HCO}, \ce{CH2OH}, and \ce{CH3O} with the radicals \ce{CH3}/\ce{CH2}, for example \ce{CH3CHO}, \ce{C2H5OH}, and \ce{CH3OCH3}. Our results also align with the observed segregation of COMs in N- and O-bearing COMs presented in \cite{JImenezSerra2016, Jimenez-Serra2021, megias_complex_2022} in the starless cores L1544, L1498, and L1517B. These observations show that O-bearing molecules and specifically \ce{CH3CHO} abundances are located toward the center of the core. On the contrary, N molecules are located at the edges of the cloud, making them more efficiently formed in earlier stages, possibly because of gas-phase nitrogen chemistry. Based on the evidence of efficient cyanation reactions \citep{Balucani1999, Vazart2015, Puzzarini2020, Jule2022} in the gas phase and the evidence collected in this work, the enhanced chemistry in CO appears to be more than plausible for the increased abundance of O-bearing COMs. This is in line with our models, where the impact of CO chemistry in the abundance of N-bearing COMs is less pronounced than that of O-bearing COMs. In the models presented in Sections \ref{sec:modelCOCH3} and \ref{sec:modelHCO}, we show abundances of \ce{CH3NH2}, \ce{CH3CN}, and \ce{NH2CHO}, but this observation also applies to other molecules such as \ce{C2H3CN} or \ce{C2H5CN}. We note that the abundance of \ce{CH3NH2} is high in the grains (Figure \ref{fig:heatmaps:12K}), but its abundance is also high in models that do not consider CO ice. Therefore, its chemistry on the CO ice is not as enhanced as in the case of O-bearing COMS. The reason for this chemical segregation is the importance of CO hydrogenation products in the promotion of COM formation. These products (\ce{HCO}, \ce{CH2OH}, \ce{CH3O}) are precursors to O-bearing COM.

\subsection{What inputs are necessary to improve the models?}

It is evident from our simulations that CO-dominated ice promotes richer chemistry than the one present on \ce{H2O} ice, which is wholly dominated by hydrogenations at low temperatures, $\lesssim$ 20 K. Considering the relatively fast diffusion of radicals on CO ice, the diffusive constraints for several debated grain routes for the formation of COMs on ice are lifted, such as \ce{NH2CHO} \citep{Noble2015,Song2016,Rimola2018, lopez-sepulcre_interstellar_2019} or \ce{CH3CHO} \citep{Enrique-Romero2016, Lamberts2019, enrique-romero_theoretical_2021}. Despite models represent a theoretical effort to characterize the phenomenon on a qualitative level, physicochemical input from different sources can increase the predictive potential of the models, especially with regard to gas-phase species and comparison with observations.

The uncertainties in the rate equation model are mostly related to the $E_{\text{diff}}$/$E_{\text{bin}}$ ratio (See Appendix \ref{sec:appendix3}). Because CO chemistry occurs in a narrow range between 8--18 K, we can easily identify which radicals are vital in the production on COMs on CO ice, most notably \ce{CH3} or \ce{HCO}, depending on $T_{\rm d}$ (see Figure \ref{fig:hopping}). Should experiments or dedicated simulations provide better constraints on $E_{\text{diff}}$/$E_{\text{bin}}$ for these species, a parameter that we encourage to search for, it will be easy to update the values derived here, seeking a better agreement between observations and models of prestellar cores. 

Although less critical than the $E_{\text{diff}}$/$E_{\text{bin}}$ ratio, two other factors could improve our models. First, to improve the description of chemical desorption and our predictions concerning the abundances of the gas phase, information on the energy transfer of chemical energy to a CO ice (compared to a \ce{H2O} ice) is required. Although intuition suggests that chemical desorption must be enhanced on CO ice, due to the lower BE of the adsorbates on the surface, it is not clear how chemical energy transfer occurs in these ices. In our models, energy dissipation is governed by $a$ (Equation \ref{eq:cd_frac}), and is assumed to be equal on CO ices and \ce{H2O} ices. The lack of knowledge about the differences in chemical desorption probabilities on different substrates makes us consider a similar desorption probability, something that can be easily improved when new investigations address this particular issue. The second input that could also slightly improve our models is the inclusion of exclusive CO chemistry, that is, reactions that can only occur on CO ice. To our knowledge, the only example to date involves the condensation of atomic carbon on CO ice \citep{Fedoseev2022, ferrero2023formation}, a reaction whose modelling is the subject of a subsequent work. When new reactions are postulated, our models can easily incorporate them.

Finally, it is vital to indicate that our models' initial conditions are tailored to produce CO-rich ice, allowing us to investigate the maximum impact of CO-dominated ice. Today, there is no clear consensus on the morphology of the CO/\ce{H2O} ice in interstellar environments, for example, whether it is layered \citep{Boogert2015} or, on the contrary, the different components of the ice are mixed, forming islands on the core of the grain \citep{Potapov2020}. Recent JWST results support the presence of pure CO ice without significant mixing with \ce{H2O} ice \citep{McClure2023}. Even assuming the creation of layered ice, it is not straightforward to confirm whether interstellar CO completely wets the surface or not \citep{noble2012thermal,Kouchi2021} or whether CO ice is crystalline or amorphous \citep{Kouchi2021, he_phase_2021}. These questions need to be answered by astronomical observations of interstellar ice. Our models coarse-grain all these unknowns via a continuous function of BE as a function of the CO surface coverage (Eq. \ref{eq:coverage}). We will continue monitoring the recent discoveries in interstellar ice structure through JWST observations and adapt our models accordingly.

\subsection{Conclusions}

We conclude this work with a series of bullet points that can be extracted from this work:

\begin{enumerate}
    \item The diffusion of relatively heavy radicals on CO-ice (\ce{CH3}, \ce{HCO}, N, O, \ce{CH2} and \ce{NH}) enhances alternative reactions to the hydrogenations on said ice. In particular, the formation of \ce{CH3CHO}, \ce{CH3OCH3}, and \ce{C2H5OH} appears to be significantly enhanced on CO ices. 
    \item The dust temperature contribution to the chemistry on CO-dominated ice is hardly predictable at first sight and needs to be considered on a per-reaction basis, although the competition between H atom diffusion and desorption versus other radicals diffusion is the main driving mechanisms for the observed changes.
    \item The enhanced COM formation on CO-dominated ice is inhibited when $\zeta$ is high (normally \textgreater 5$\times$10$^{-17}$ s$^{-1}$) \textbf{and $T_{\rm d}$ is low}. High ionization rates enhance H atom abundances and thus hydrogenation of radicals. Molecules that can be formed by hydrogenation of simpler precursors like CO, \ce{C2N} or \ce{C2H2}  (yielding \ce{CH3OH}, \ce{CH3CN}, \ce{C2H4}) do not see an appreciable increase in their abundances due to competition between heavy radical diffusion and H diffusion on CO ices.
    \item There are two regimes for the enhancement of chemistry on CO ice. The first occurs for T$_{\text{d}}\leq$12 K, where we observe an enhancement of the chemistry initiated by \ce{CH3}, \ce{CH2}, O or N. A second regime for enhanced COMs formation is revealed above $\geq$ 14 K, initiated by the chemistry of HCO and NH. This second regime is less impactful due to the strong competition with hydrogenation. Still, the increase in abundance of COMs like \ce{NH2CHO}, \ce{HCOOCH3}, \ce{OHCCHO}, \ce{HOCH2CHO} and \ce{NH2CH2COOH} on CO dominated ices is notable, at low $\zeta$.
    \item The desorption of species from CO-dominated ices is also promoted under several conditions, such as an overall small binding energy of the desorbed molecule or a greater COM formation compared to the \ce{H2O}-dominated ices. However, extracting quantitative gas abundances becomes even more complicated when considering gas-phase chemistry in addition to the uncertainties in the model.
    \item Our models can be enhanced once several unknowns are resolved from experiments or quantum chemical calculations. For example, we highlight the determination of $E_{\text{diff}}$/$E_{\text{bin}}$ on CO ices in comparison with \ce{H2O} ices, the rate of chemical energy transfer for chemical desorption on CO ices, and the study of reactions taking place exclusively on CO ice. 
\end{enumerate}

The results presented in this paper serve as a step towards improving the description of the chemistry on interstellar dust grains and complement other contemporary advances, such as nondiffusive chemistry. We will continue improving our models, based on new observational evidence and theoretical advances. An example of these new techniques includes the recent approach presented in \cite{Garrod2022} of considering different binding sites on a substrate based on a probabilistic distribution of binding partners.

Our results are timely in the context of JWST observations. We will also continue expanding the quantum chemical dataset in search of binding energy outliers and adapting our methodology to simulate realistic pre-stellar cores, including the physical evolution of the objects in the astrochemical models.

\begin{acknowledgements}
We thank Dr. Joan Enrique-Romero for his insightful comments. G.M. thanks the Japan Society for the Promotion of Science (JSPS International Fellow P22013, and Grant-in-aid JP22F22013) for its support. \textbf{G.M also acknowledges support from the grant RYC2022-035442-I funded by MCIN/AEI/10.13039/501100011033 and ESF+}. The authors acknowledge support by the state of Baden-Württemberg through bwHPC and the German Research Foundation (DFG) through grant no INST 40/575-1 FUGG (JUSTUS 2 cluster) and the Research Center for Computational Science, Okazaki, Japan (Projects: 22-IMS-C301, 23-IMS-C128). KF acknowledges support from JSPS KAKENHI grant Numbers 20H05847 and 21K13967. Y.A. acknowledges support by Grant-in-Aid for Scientific Research (S) 18H05222 and Grant-in-Aid for Transformative Research Areas (A) grant Nos. 20H05847.
\end{acknowledgements}

%
%

\bibliographystyle{aa}
\bibliography{sample.bib}

\begin{appendix}
\section{Quantum Chemical Determination of $\epsilon_{\text{CO}}$ for \ce{CH3}, \ce{HCO}, N, O, \ce{CH2}, NH and \ce{H2}.} \label{sec:appendix1}

\begin{figure}
  \centering        
  \includegraphics[width=\linewidth]{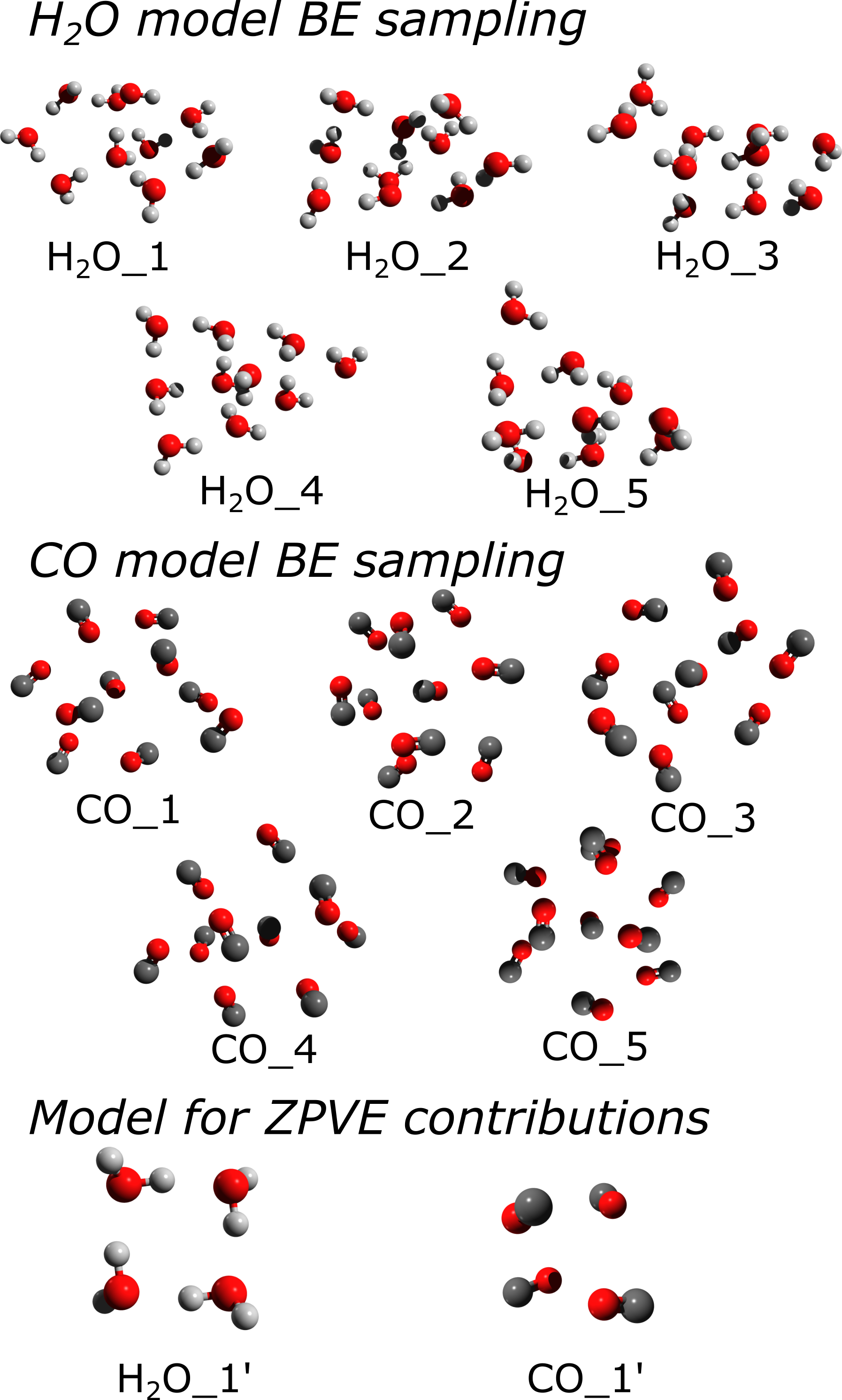} \\
  \caption{\textbf{Structures of the clusters used for the sampling of BE and BE(0) ZPVE corrections in this appendix. Further molecular geometries can be provided on request.}}
  \label{fig:revision}
\end{figure}

As discussed in the main text, our chemical models approximate BE on pure and mixed CO ice through the scaling of BE on water (Equation \ref{eq:scale}, $\epsilon$). The reason for this is the absence of data on the former two with respect to the latter, due to a very high computational cost of computing binding energy distributions on realistic CO ice analogues. Furthermore, because BE is varied depending on $\theta_{\text{CO}}$ at each time step, we need a formulation applicable to intermediate situations that do not contain pure CO ices or \ce{H2O} ices (Equation \ref{eq:coverage}). We use quantum-chemical calculations to derive $\epsilon_{CO}$. For CO ice clusters, we follow a similar scheme as in \citep{Ferrari2023}, creating 5 clusters with 10 CO molecules each and sampling 8 adsorption positions for each cluster, for a total of 40 adsorption geometries per adsorbate+cluster combination. \textbf{We chose to use several clusters and different adsorption geometries per cluster to account for a wide variety of binding sites for our reduced-size cluster models.} For \ce{H2O} we repeat the same sampling for the same number of molecules. \textbf{All cluster geometries are shown in Figure \ref{fig:revision}.} We then place the adsorbates (\ce{CH3}, \ce{HCO}, N, O, \ce{CH2}, NH, \textbf{and \ce{H2}}) randomly in equispaced points on the sphere surrounding the clusters, projecting the adsorbates to ensure a minimum distance of $\sim$3~$\AA$ between the center of mass of the adsorbate and the nearest neighbor on the cluster. Binding energies are then computed as the difference between the energy of the complex adsorbate+cluster and the sum of the separated components (adsorbate + cluster). The $\omega$B97 family of functionals showed excellent performance in the study of binding energies, both on CO and \ce{H2O} ice \citep{ferrero2023formation, Hendrix2023, Ferrari2023}. Therefore, for this work we used $\omega$B97M-D4/def2-TZVP \citep{Mardirossian2016, Weigend2005, Caldeweyher2019, najibi_dft_2020}. All calculations shown in this Appendix are performed using \textsc{ORCA 5.0.4} \citep{Neese2020}.

\begin{figure}
    \centering        
    \includegraphics[width=\linewidth]{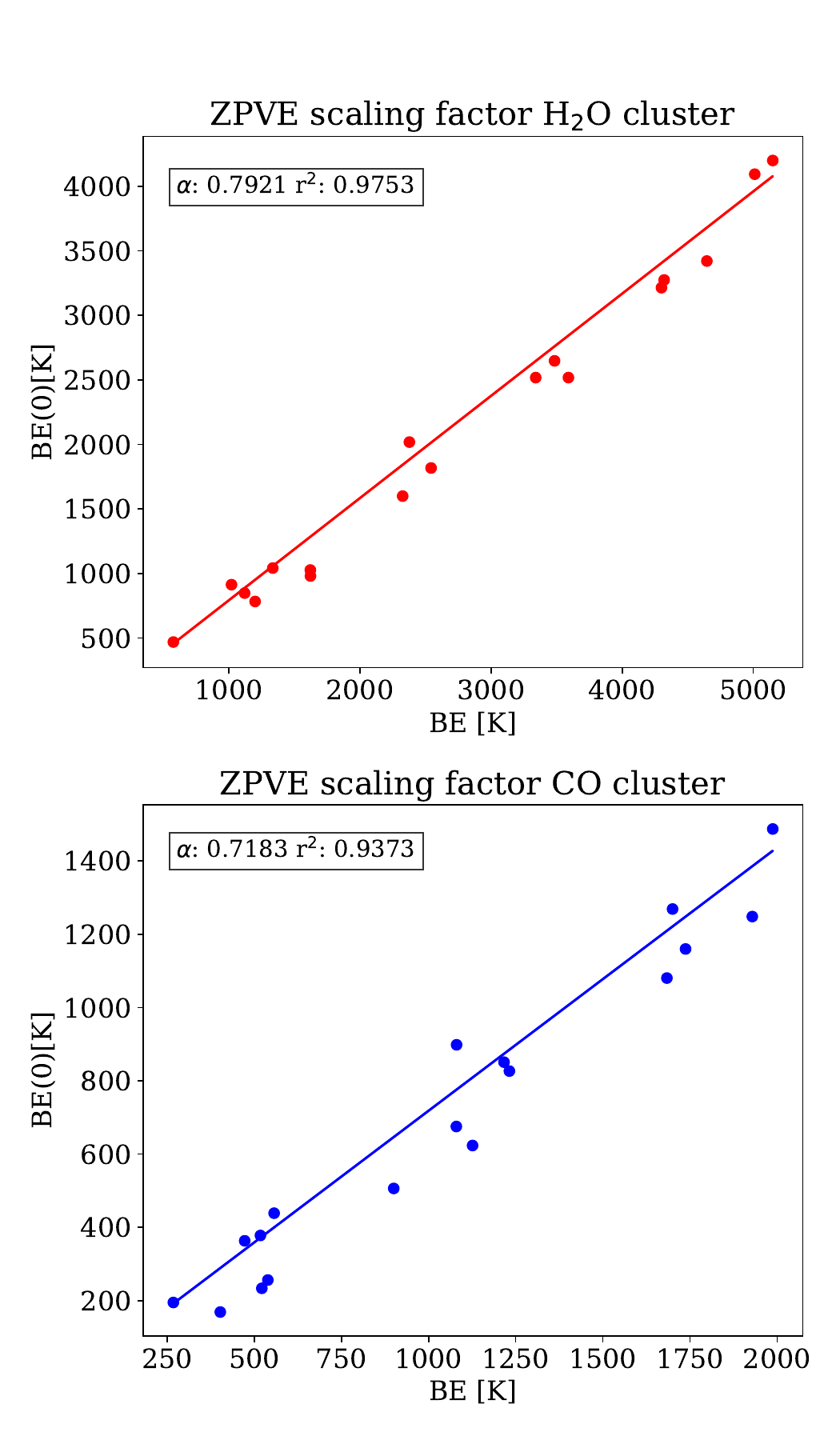} \\
    \caption{Correlation of binding energies including zero point energy contributions, BE(0), and neglecting zero point energy corrections, BE.}
    \label{fig:calibration}
\end{figure}

Zero-point energy corrections to the binding energy are introduced via a scaling factor, because of the high computational cost associated with deriving the molecular Hessian of every sampling point. We obtain the proportionality factor, $\alpha$, using the same theoretical method ($\omega$B97M-D4/def2-TZVP) and connecting non-ZPVE inclusive and ZPVE inclusive BE using a model system containing 4 molecules (either \ce{H2O} or CO) and an adsorbate test set consisting of \textbf{\ce{CH3}, \ce{HCO}, N, O, \ce{CH2}, NH, \ce{CH2OH}, \ce{CH3OH}, \ce{CH4}, CO, \ce{CO2}, \ce{H2CO}, \ce{H2O}, \ce{H3CO}, \ce{N2}, \ce{NH2}, \ce{NH3}, and OH.} This approach has been successfully applied in computational works dedicated to the determination of binding energies of collections of molecules \citep{Ferrero2020, bovolenta_binding_2022, perrero_binding_2022}. The results of this study are shown in Figure \ref{fig:calibration} \textbf{and the models used for the ice substrate are shown in Figure \ref{fig:revision}}. From the graph, we observe a correlation between the non-ZPVE corrected BE (BE) and the ZPVE corrected, BE(0), with a correlation coefficient in line with other works in the literature. Then the binding energies used to derive $\epsilon_{CO}$ are obtained as:

\begin{align} \label{eq:be}
    \text{ BE(0, \ce{H2O})} & = 0.7921\times\text{BE(\ce{H2O})} \\
    \text{ BE(0, \ce{CO})} & = 0.7183\times\text{BE(\ce{CO})}.
\end{align}

\noindent We observe a higher influence of ZPVE on CO ices than in \ce{H2O} ices, which results in a lower value of $\epsilon_{CO}$, further facilitating diffusive chemistry. We note that even though in this appendix we differentiate between BE and BE(0), in the main text BE refers to BE(0) throughout. 

\begin{figure}
    \centering        
    \includegraphics[width=\linewidth]{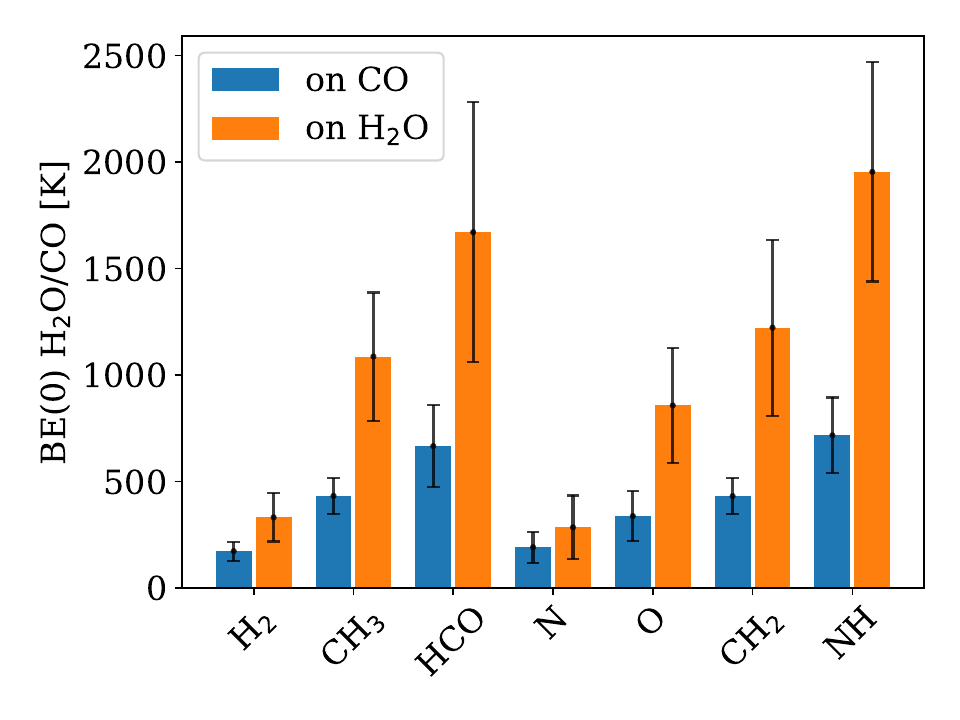} \\
    \caption{\textbf{Average of the binding energies obtained using a 10 molecule cluster for the key adsorbates considered in this work. In blue, binding energies on CO clusters, in orange, binding energies on \ce{H2O}. Error bars represent standard deviation.}}
    \label{fig:h2ovsco}
\end{figure}

Finally, after calculating BE(0) using the 40 adsorption site sampling in the 10 CO/\ce{H2O} cluster and $\alpha$, we can extract a representative binding energy as the average BE(0) considering the different adsorption sites. The results of this analysis, along with standard deviations, are collated in Figure \ref{fig:h2ovsco}. It is worth mentioning that higher deviation of BE is found for \ce{H2O} clusters, due to the variability in H-bond topologies in the different binding sites. The BE(0) ratios shown in Figure \ref{fig:h2ovsco} are the ones presented in Table \ref{tab:epsilon} of the main text. 

\textbf{The values in Table \ref{tab:epsilon} of the main manuscript and Figure \ref{fig:h2ovsco} show that, in general, $\epsilon_{\ce{CO}}$  lower than 0.40 describe well the binding of the adsorbates on CO to \ce{H2O}. This is in good agreement with \citet{mondal_is_2021}, that reports 0.35--0.40 for molecules with permanent dipole moment or significant polarizability. Both our calculations and \citet{mondal_is_2021} show that molecules that form H-bonds, dipole-dipole, or effective dipole-induced dipole (as in \ce{CH3}) interactions will have $\epsilon_{\ce{CO}}$ below 0.4. Among the molecules that we considered in Table \ref{tab:epsilon}, exceptions are N, CO, and H (and also \ce{H2}, not shown in the table); they are apolar, small, and thus hardly polarizable. In the absence of a permanent dipole to form dipole-dipole interactions, adsorbate surface interactions are driven by weaker forces, depending on the polarizability of the adsorbate. This allows us to rationalize the unity value found for H atoms \citep{Fuchs2009, Hama2012, Kimura2018}, because the interactions affecting the H atom, the smallest possible chemical adsorbate, are essentially the same on \ce{H2O} and \ce{CO} ices. Likewise, a similar case can be found for the N atom, but here the species is larger, and a certain degree of polarizability is likely admitted. In fact, the phosphorous atom (P) \citet{sil_chemical_2021} finds an $\epsilon_{\ce{CO}}$ of 0.53, using a monomer model, which follows the polarizability trend that we discussed in this paragraph. Not surprisingly, different P-bearing molecules have different $\epsilon_{\ce{CO}}$ depending on the interactions at hand. Because P-bearing molecules are out of the scope of this modeling effort we refer the reader to the work of \citet{sil_chemical_2021}. Continuing with the adsorbates with large  $\epsilon_{\ce{CO}}$ we have \ce{H2}, that according to \citet{Molpeceres2020} and \citet{Molpeceres2021a}, should have $\epsilon_{\ce{CO}}$ of around 0.5, and CO, that according to \citep{Bisschop2006, Fuchs2009, Furuya2018, Ferrari2023} should have values above 0.60 (0.66). The difference between these apolar, hardly polarizable, small molecules stems most likely from their quadrupole moments. Carbon monoxide having a larger quadrupole moment is reflected by a stronger binding with CO ice, in contrast with \ce{H2} whose quadrupole moment is significantly lower. }

\section{Chemical models varying $\epsilon_{\ce{CO}}$ for the majority of radicals} \label{sec:appendix2}

\begin{figure}
    \centering
    \includegraphics[width=\linewidth]{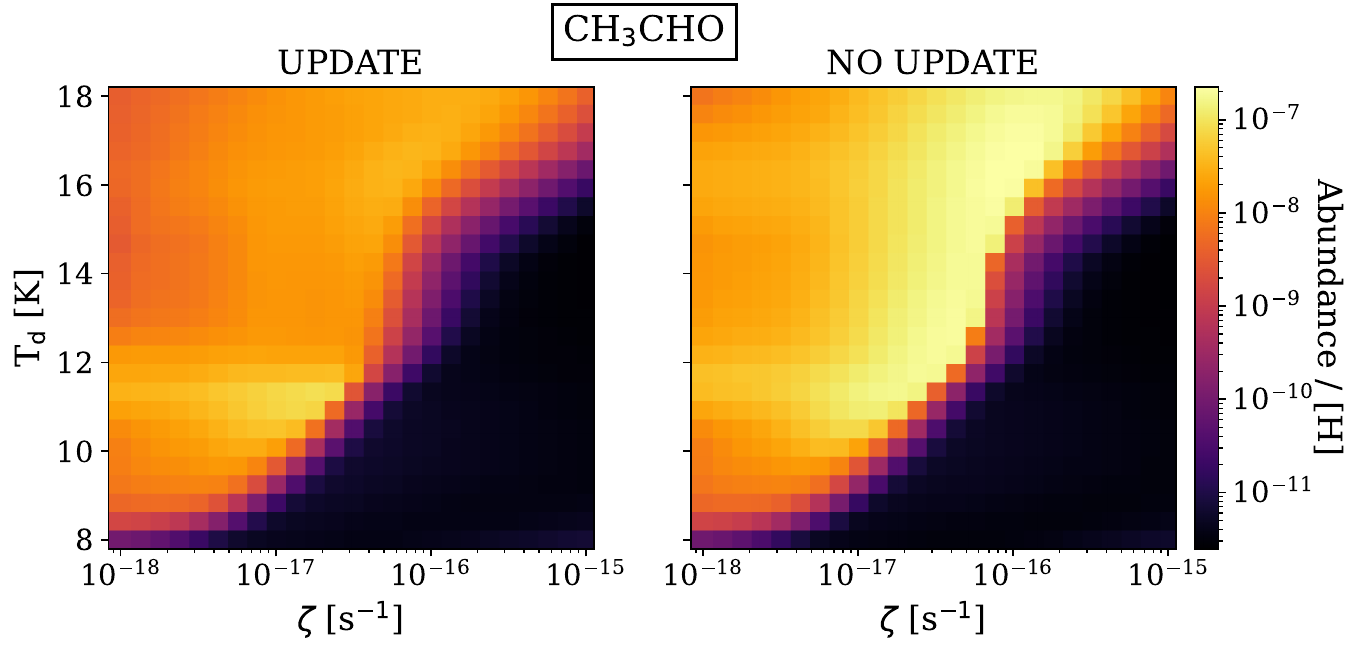} \\
    \includegraphics[width=\linewidth]{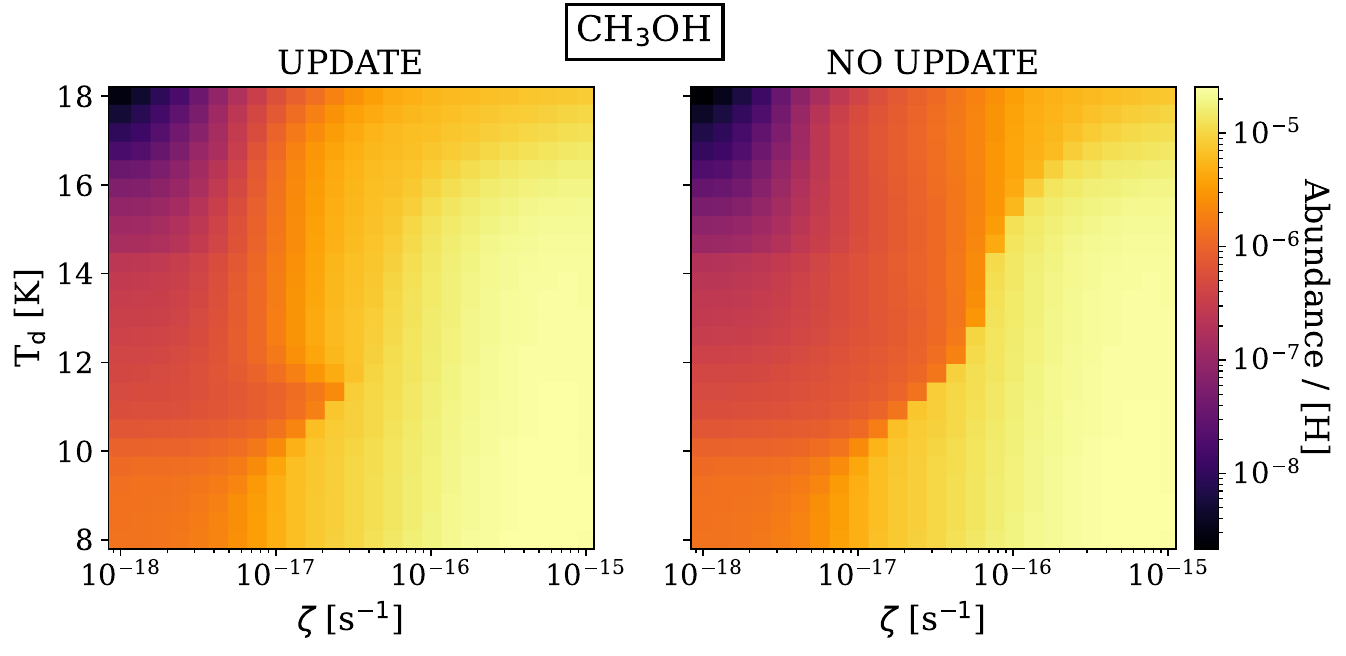} \\
    \includegraphics[width=\linewidth]{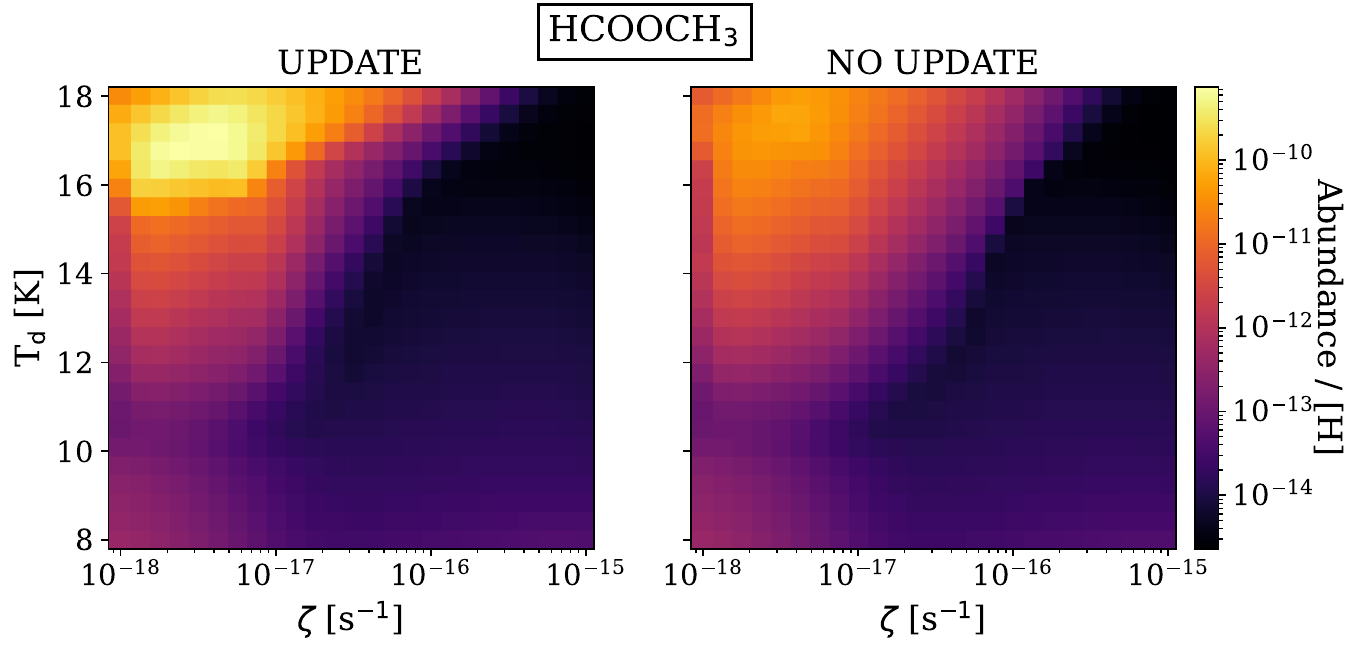} \\
    \caption{\textbf{Comparison of abundances for a reduced grid of models considering an update of BE through $\epsilon_{\text{CO}}$ of 0.66 (labeled ``UPDATE") or 1.00 (labelled ``NO UPDATE") for all species labeled as ``Rest of surface species" in \ref{tab:epsilon}. The color scheme is chosen deliberately different to the main text.}}
    \label{fig:heatmaps:comparison1}
\end{figure}

Our production models update all surface species not included in Table \ref{tab:epsilon} with a constant scaling using $\epsilon_\text{CO}$=0.66. While this selection is made to ensure that all species in the reaction network can have their binding energies updated, the selection of $\epsilon_\text{CO}$=0.66 is based on the ratio found for CO \citep[][and references therein]{Furuya2018} and may be different for other species. To confirm that such a choice does not significantly alter our results, we ran additional models using a reduced grid for T$_{d}$ and $\zeta$ (e.g. 25x25), considering $\epsilon_\text{CO}$ = 0.66 for all adsorbates not explicitly mentioned in Table \ref{tab:epsilon} and $\epsilon_\text{CO}$=1.00, e.g. not updating BE for those molecules. The results are visually shown for \ce{CH3CHO}, \ce{CH3OH} and \ce{HCOOCH3} in Figure \ref{fig:heatmaps:comparison1} where the color scheme is selected differently from all the other heatmaps to highlight the comparison. A visual inspection of Figure \ref{fig:heatmaps:comparison1} shows that although there are differences between the models, these are only minor, as shown by the similar profiles in the heatmaps, especially compared with a model without a BE update (See Figures \ref{fig:heatmaps:12K} and \ref{fig:heatmaps:14K}). These small changes demonstrate that the diffusion of the selected species in Table \ref{tab:epsilon} is the main reason behind the chemistry enhancement on CO ice.  



\section{The effect of $E_{\text{diff}}$/$E_{\text{bin}}$} \label{sec:appendix3}

\begin{figure}
    \centering
    \includegraphics[width=\linewidth]{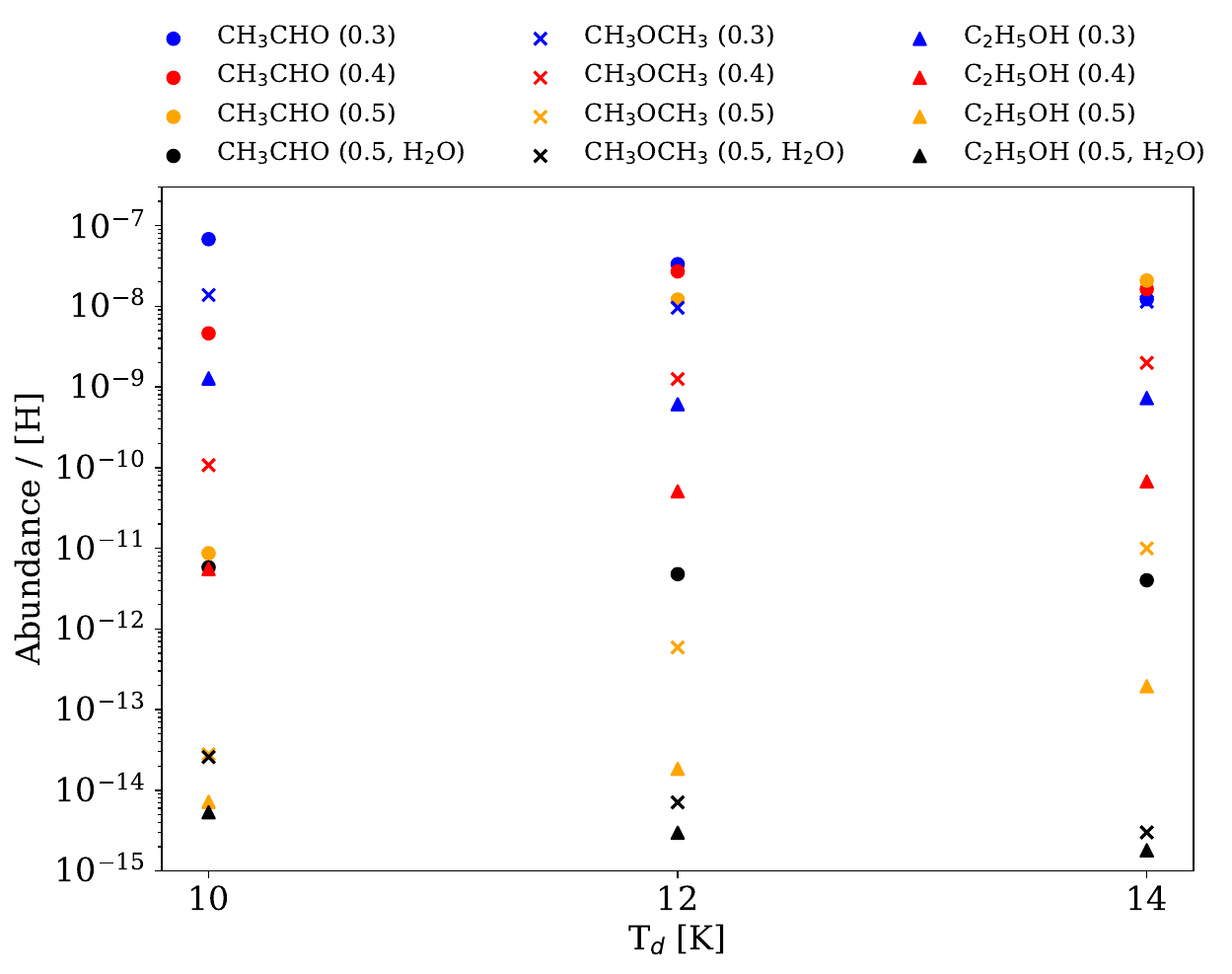} \\

    \caption{\textbf{Steady state abundances (i.e 1$\times$10$^{6}$) yr for \ce{CH3CHO}, \ce{CH3OCH3}, and \ce{C2H5OH} for fixed T$_{d}$. and $\zeta$=1.3$\times$10$^{-17}$ s$^{-1}$. At each T$_{d}$ we vary the $E_{\text{diff}}$/$E_{\text{bin}}$ between 0.3--0.5 for the models with updated BE on CO ice. For comparison, we also show the models for $E_{\text{diff}}$/$E_{\text{bin}}$=0.5 without the update of BE (labeled 0.5, \ce{H2O} in the graphic's legend).} }
    \label{fig:heatmaps:ediff}
\end{figure}

As we discussed in the main text, the biggest unknown of our models is the diffusion vs. desorption ratio ($E_{\text{diff}}$/$E_{\text{bin}}$), which is the main parameter determining the mobility of chemical species on ices. To show how $E_{\text{diff}}$/$E_{\text{bin}}$ affects our results, we run models for fixed T$_{d}$ and $\zeta$ for different values $E_{\text{diff}}$/$E_{\text{bin}}$ in comparison to a model without updating $\epsilon_{\text{CO}}$ and $E_{\text{diff}}$/$E_{\text{bin}}$=0.5. In the main text, the fiducial model considers $E_{\text{diff}}$/$E_{\text{bin}}$=0.4, but since in this appendix the maximum $E_{\text{diff}}$/$E_{\text{bin}}$ is 0.5 the comparison needs to be held taking this value as a reference. The steady-state abundances for each of these conditions at 10, 12, and 14 K are shown in Figure \ref{fig:heatmaps:ediff} for \ce{CH3CHO}, \ce{CH3OCH3} and \ce{C2H5OH}, which are much affected by the update of BE in section \ref{sec:modelCOCH3}. The effect of $E_{\text{diff}}$/$E_{\text{bin}}$ shown in Figure \ref{fig:heatmaps:ediff} is evident, especially at 10 K where the steady-state abundances can vary by several orders of magnitude. This effect is diminished at higher T$_{d}$. While better constraints on $E_{\text{diff}}$/$E_{\text{bin}}$ on CO ices would surely allow for a more quantitative prediction of ice and gas phase abundances, it is clear from Figure \ref{fig:heatmaps:ediff} that models that update the BE on CO ices yield systematically higher abundances than models without such an update (Shown by black lines in Figure \ref{fig:heatmaps:ediff}). Such a conclusion demonstrates that our hypothesis in this work, that is CO ices promote COMs chemistry, holds for different $E_{\text{diff}}$/$E_{\text{bin}}$ values.

\end{appendix}

\end{document}